\documentclass{ptephy_v1}

\preprintnumber{TIT/HEP-646} 

\usepackage{bm}
\usepackage{amsmath}
\usepackage{amsfonts}

\newcommand{\II}{\mathbb{I}}

\newcommand{\ZZ}{\mathbb{Z}}
\newcommand{\RR}{\mathbb{R}}
\newcommand{\ol}{\overline}
\newcommand{\wh}{\widehat}
\newcommand{\wt}{\widetilde}
\newcommand{\CP}{{\mathbb{CP}}}
\def\Pexp{\mathop{\rm Pexp}\nolimits}
\def\tr{\mathop{\rm tr}\nolimits}
\def\sign{\mathop{\rm sign}\nolimits}
\def\rot{\mathop{\rm rot}\nolimits}
\def\diag{\mathop{\rm diag}\nolimits}

\newcommand{\rap}[2]
{\setbox1=\hbox{#1}%
\setbox2=\hbox to\wd1{\hss #2\hss}%
\mbox{\rlap{\box1}\box2}}
\newcommand{\sla}[1]{\rap{$#1$}{$\backslash$}}

\begin{document}

\title{Index calculation by means of harmonic expansion}


\author{Yosuke Imamura}
\affil{Department of Physics, Tokyo Institute of Technology, Tokyo 152-8551, Japan\email{imamura@phys.titech.ac.jp}}


\begin{abstract}%
We review derivation of superconformal indices
by means of
supersymmetric localization and
spherical harmonic expansion
for 3d ${\cal N}=2$, 4d ${\cal N}=1$, and 6d ${\cal N}=(1,0)$ supersymmetric gauge theories.
We demonstrate calculation of indices for vector multiplets
in each dimensions
by analysing energy eigenmodes in ${\bm S}^p\times\RR$.
For the 6d index we consider the perturbative contribution only.
We put focus on technical details of harmonic expansion rather than
physical applications.
\end{abstract}

\subjectindex{B00, B16}

\maketitle

\section{Introduction}\label{intro.sec}

Indices are special type of partition functions
in supersymmetric theories
defined in such a way that the contributions of
bosonic and fermionic states
partially cancel each other and only modes satisfying
some BPS conditions make non-trivial contributions.
Thanks to the supersymmetry, indices are protected from
uncontrollable quantum corrections,
and it is possible to calculate them exactly
even in strongly coupled theories.
They are powerful tools to analyze dynamics
of supersymmetric theories
and have been used to test various dualities.

There are variety of indices.
The simplest one is the Witten index \cite{Witten:1982df},
which is defined by
\begin{align}
I_W=\tr[(-1)^Fe^{-\beta H}],
\end{align}
where $H$ and $F$ are the Hamiltonian and the fermion number, respectively.
The trace is taken over all gauge invariant states.
This gives an integer, which is the difference between the numbers of
bosonic and fermionic vacua.
To calculate $I_W$ the system is put in a torus to avoid IR divergence.

We can generalize $I_W$ by replacing the background space
with another compact manifold $M$ (or a non-compact manifold with appropriate boundary conditions).
If we take $M$ to be ${\bm S}^p$, the index gives BPS spectrum of the theory in ${\bm S}^p\times\RR$.
We can map the background to $\RR^{p+1}$ by a Weyl transformation,
and for superconformal theories
the index can be associated with local BPS operators
via the state-operator correspondence.
This kind of indices, which we will focus on in this paper, are called superconformal indices
\cite{Romelsberger:2005eg,Kinney:2005ej,Bhattacharya:2008zy}.
The purpose of this paper is to demonstrate derivation of
the superconformal indices for 3d, 4d, and 6d supersymmetric theories using harmonic expansion.

For a given $p$-dimensional manifold $M$,
the index is defined and calculated as follows.
We consider a $p+1$ dimensional supersymmetric field theory
defined on $M\times\RR$.
We suppose the theory has $k$ commuting bosonic conserved charges including
the Hamiltonian $H$ and other $k-1$ charges $F_i$ ($i=1,\ldots,k-1$).
Let $Q$ be a supercharge,
and $\ol Q$ be another supercharge such that $\{Q,\ol Q\}=H+c_iF_i$.
The index is defined by
\begin{align}
I(e^{-\beta},e^{-\gamma_i})=\tr[(-1)^Fe^{-\beta H-\gamma_iF_i}].
\label{indexdef}
\end{align}
The operator $\beta H+\gamma_iF_i$ in the exponent must commute with $Q$.
Therefore, only $k-1$ variables in $(\beta,\gamma_i)$ are independent.
It is easy to show that the deformation of the Hamiltonian by a $Q$-exact term
\begin{align}
H\rightarrow H'=H+t\{Q,V\}
\label{defham}
\end{align}
does not change (\ref{indexdef}).
If we take $V=\ol Q$, the deformation is equivalent to the shift of parameters
$(\beta,\gamma_i)\rightarrow(\beta,\gamma_i)+t(1,c_i)$.
This means that the index depends on the parameters through $\gamma_i-\beta c_i$,
and the index is a function of $k-2$ variables.

If we choose $V$ so that the deformation term contains quadratic terms of fields,
the theory becomes weakly coupled when we take $t\rightarrow\infty$ limit.
This enables us to
calculate the index exactly from the information of
eigenmodes of the deformed Hamiltonian in the weak coupling limit.

Let $\Phi$ be fields in the theory.
We expand the solution of the free field equations
by eigenmodes of charges $(H,F_i,t_I)$
labeled by $n$,
where $t_I$ are Cartan generators of the gauge group.
Let $(\omega_n,f_{i,n},t_{I,n})$ be the eigenvalues of a mode $n$.
The solution is given by
\begin{align}
\Phi(\tau,x)=\sum_nc_n e^{-\omega_n\tau}Y_n(x),
\end{align}
where $\tau$ is a Euclidean time and $x=(x^1,\ldots,x^p)$ are
coordinates in $M$.
The frequencies $\omega_n$ are always non-zero
in the theories we consider in the following sections
except for gauge degrees of freedom,
which have to be dealt with separately.

The spectrum of multi-particle states in a free theory is
uniquely determined by the spectrum of single-particle states.
It is convenient to define the single-particle index
in a similar way to
(\ref{indexdef}) by
\begin{align}
I_{\rm sp}(e^{-\beta},e^{-\gamma_i},e^{ia_I})
&=\tr[(-1)^F
e^{-\beta H-\gamma_iF_i+ia_It_I}]
\nonumber\\
&=\sum_n (-)^F
\left(e^{-\beta\omega_n-\gamma_if_{i,n}+ia_It_{I,n}}\right)^{\sign\omega_n},
\label{ispdef}
\end{align}
where the trace is taken over all single-particle states
including charged states.
The exponent, $\sign\omega_n$, represents the fact that
negative frequency modes correspond to anti-particles,
which carry opposite quantum numbers to those of particles corresponding to
positive frequency modes.
If the field is complex we have to take account of both
positive and negative frequency modes,
while for a real field we need to include
only positive (or negative) frequency modes.
The index for multi-particle states is given by
\begin{align}
I_{\rm mp}(e^{-\beta},e^{-\gamma_i},e^{ia_I})
=\Pexp I_{sp}(e^{-\beta},e^{-\gamma_i},e^{ia_I}),
\label{impdef}
\end{align}
where $\Pexp$ is the plethystic exponential defined by
\begin{align}
\Pexp f(x,y,\ldots)=\exp\left(\sum_{n=1}^\infty\frac{1}{n}f(x^n,y^n,\ldots)\right).
\end{align}
This multi-particle index includes the contribution of
charged states as well as gauge invariant states.
To obtain the index for physical states,
we need to pick up the contribution of gauge invariant states
by integrating over $a_I$.
To determine the measure of this integral let us rewrite
(\ref{indexdef}) in the path integral form
\begin{align}
I(e^{-\beta},e^{-\gamma_i})=\int [d\Phi]e^{-S-S'},
\end{align}
where $S$ is the supersymmetric action of the theory defined on the background
$M\times{\bm S}^1$, and $S'=t\delta_QV$ is a $Q$-exact deformation term
corresponding to the deformation in (\ref{defham}).
The parameters $\beta$ and $\gamma_i$ in (\ref{defham}) are taken in the path integral
formulation as the period of ${\bm S}^1$ and the Wilson lines
associated with the global symmetries, respectively.
In the weak coupling limit the saddle point approximation
gives the exact answer.
For a gauge theory we have to fix the gauge symmetry and
carefully take the associated ghost factor into account.
Let $\alpha$ be weights in the adjoint representation of the gauge group ${\cal G}$
including both roots and Cartan generators,
and $t_\alpha$ be the corresponding generators.
Let $A$ be a fluctuation of the gauge potential around a saddle point.
It belongs to the adjoint representation and is expanded as
\begin{align}
A=\sum_\alpha A^\alpha t_\alpha
=\sum_{\alpha\neq0} A^\alpha t_\alpha
+\sum_I A^It_I,
\end{align}
where $\alpha\neq0$ means the sum is taken over only roots,
and $I$ labels the Cartan generators.
We divide the gauge field in $M\times{\bm S}^1$ into
two parts, the $x$-independent part $A^0$
and the $x$-dependent part $A'\equiv A-A^0$.
The gauge fixing for $A'$ will be done in the
following sections.
We here focus only on $A^0=A_\tau^0(\tau)d\tau$.
The path-integral of $A^0$ essentially gives the integral over
the Wilson line $u={\rm P}\exp\int_{\bm S}^1 A^0\in{\cal G}$.
We take the static gauge
\begin{align}
A_\tau(\tau)=\frac{1}{\beta}\sum_Ia_It_I,
\end{align}
where $a_I$ are $\tau$-independent constant,
which is identified with the chemical potentials for $t_I$ in (\ref{ispdef}).
In this gauge
the integral over $u$
can be rewritten as
\begin{align}
\frac{1}{{\rm Vol}\ {\cal G}}\int du I_{\rm mp}
=\frac{1}{|W|}
\int\prod_{I=1}^r\frac{da_I}{2\pi}
\prod_{\alpha\neq0}
\left(1-e^{i\alpha(a)}\right) I_{\rm mp},
\label{qmfix}
\end{align}
where $r={\rm rank}\ {\cal G}$ and $|W|$ is the number of elements of the Weyl group.
$\alpha(a)$ is defined by
$[a,t_\alpha]=\alpha(a)t_\alpha$
for a Cartan element $a$ of the gauge algebra.
As well as the non-zero mode contribution (\ref{impdef}),
the measure factor in (\ref{qmfix}) is also written in the form of the plethystic exponential
of the single-particle index
\begin{align}
I_{\rm sp}^{({\rm gh})}(e^{ia_I})
=-\sum_{\alpha\neq0} e^{i\alpha(a)}
=\sum_{\alpha} e^{i\alpha(a)}
(\delta_{\alpha,0}-1).
\label{ispghost}
\end{align}
We can interpret this as the contribution of ghost constant modes.

After all, we obtain the following final expression for the
index
\begin{align}
I
=\frac{1}{|W|}
\int\prod_{I=1}^r\frac{da_I}{2\pi}
\Pexp I_{\rm sp}^{({\rm tot})},\quad
I_{\rm sp}^{({\rm tot})}=
I_{\rm sp}+I_{\rm sp}^{({\rm gh})}.
\end{align}

In the rest of the paper we calculate
the single particle index $I_{\rm sp}$ for vector multiplets
in 3, 4, and 6 dimensional theories.
For 6d theories
we only take account of the perturbative contribution.

\section{Spherical Harmonics}\label{harm.sec}
In this section, we define spherical harmonics
on ${\bm S}^p$
in preparation for the analysis in the following sections.

\subsection{Local frame on ${\bm S}^p$}
Let ${\bm f}_a$ ($a=-1,0,1,\ldots,p-1$) be $p+1$
unit vectors that form
an ortho-normal basis in $\RR^{p+1}$.
We denote the position vector by ${\bm y}$,
and the ortho-normal coordinates $y^a$ are defined by ${\bm y}={\bm f}_ay^a$,
or, equivalently, $y^a={\bm f}_a\cdot{\bm y}$.
The unit sphere ${\bm S}^p$ is defined by
$y_ay_a\equiv {\bm y}\cdot{\bm y}=1$.

Let $\wh T_{ab}$ be the generators of the rotation group $G\equiv SO(p+1)$
that act on ${\bm f}_a$ as
\begin{align}
\wh T_{ab}{\bm f}_c={\bm f}_a\delta_{bc}-{\bm f}_b\delta_{ac}
={\bm f}_d\rho_{dc}^V(T_{ab}),
\end{align}
where $\rho^V_{cd}(T_{ab}):=\delta_{ac}\delta_{bd}-\delta_{ad}\delta_{bc}$
are representation matrices for the vector representation.
For an anti-symmetric tensor $\lambda_{ab}$
we denote the corresponding $G$ generator $\frac{1}{2}\lambda_{ab}\wh T_{ab}$
by $\wh\lambda$.
It acts on a vector ${\bm v}={\bm f}_av^a$ as
\begin{align}
\wh\lambda{\bm v}
=\frac{1}{2}\lambda_{ab}\wh T_{ab}({\bm f}_cv^c)
={\bm f}_a\lambda_{ab}v^b.
\end{align}

${\bm S}^p$ can be given as the coset $G/H$,
where $H=SO(p)$ is the subgroup of $G$
that does not move a specific vector ${\bm n}\in {\bm S}^p$.
We choose $\bm n$ to be ${\bm n}={\bm f}_{-1}$.
${\bm S}^p=G/H$ means that there exists a projection map $\pi:G\rightarrow{\bm S}^p$
defined by
\begin{align}
\pi: g\in G\rightarrow
{\bm y}
=g{\bm n}\in{\bm S}^p.
\end{align}
In other words, $G$ is an $H$ fibration over ${\bm S}^p$.
This fiber bundle is called the frame bundle of ${\bm S}^p$.

Let $g$ be a section of the frame bundle.
Namely, $g$ is a map from ${\bm S}^p$ to $G$
satisfying ${\bm y}=g({\bm y}){\bm n}$.
With this section
we can define a local basis
${\bm\xi}_i^{({\bm y})}$
at every point ${\bm y}\in{\bm S}^p$ by
\begin{align}
{\bm\xi}^{({\bm y})}_i=g({\bm y}){\bm f}_i
\quad(i=0,1,\ldots,p-1).
\end{align}
We call $g({\bm y})$ a frame section.

The vielbein $1$-form $e^i$
and the spin connection $1$-form $\omega_{ij}$
are defined by the relations
\begin{align}
d{\bm y}={\bm\xi}^{({\bm y})}_ie^i,\quad
d{\bm\xi}^{({\bm y})}_i=
{\bm\xi}^{({\bm y})}_j\omega_{ji}+\kappa_i{\bm y},
\label{dydxi}
\end{align}
where $\kappa_i$ is the
extrinsic curvature $1$-form,
which we are not interested in.
Two equations in (\ref{dydxi}) are equivalent to
\begin{align}
g^{-1}dg=\wh e+\wh \omega,\quad
\wh e\equiv e^i\wh T_{i(-1)},\quad
\wh\omega\equiv\frac{1}{2}\omega_{ij}\wh T_{ij},
\label{ewformula}
\end{align}
and $\kappa_i=-e^i$.
(\ref{ewformula}) is used in the following to rewrite
covariant derivatives of harmonics in an algebraic form.

A change of the frame section by
$g'({\bm y})=g({\bm y})h({\bm y})$,
$h({\bm y})\in H$
reproduces 
the local frame rotation
\begin{align}
\wh e'
=h^{-1}\wh e h,\quad
\wh\omega'
=h^{-1}(\wh\omega+d)h.
\end{align}

\subsection{Spherical harmonics}
A scalar harmonic $Y^0:{\bm S}^p\rightarrow\RR$ with angular momentum $\ell$ is given
by a homogeneous polynomial of the orthonormal coordinates
$y^a={\bm y}\cdot{\bm f}_a$ of order $\ell$.
\begin{align}
Y^0({\bm y})=c_{a_1\cdots a_\ell}y^{a_1}\cdots y^{a_\ell}.
\label{scalary}
\end{align}
The coefficients $c_{a_1\cdots a_\ell}$ are components of a totally symmetric
traceless tensor satisfying
$c_{bba_3\cdots a_\ell}=0$.
We can rewrite (\ref{scalary}) as
\begin{align}
Y^0({\bm y})
&=
c_{a_1\cdots a_\ell}
({\bm y}\otimes\cdots\otimes{\bm y})
\cdot
({\bm f}_{a_1}\otimes\cdots\otimes{\bm f}_{a_\ell})
\nonumber\\
&=
g({\bm y})({\bm n}\otimes\cdots\otimes {\bm n})
\cdot
c_{a_1\cdots a_\ell}
({\bm f}_{a_1}\otimes\cdots\otimes{\bm f}_{a_\ell})
\nonumber\\
&=
g({\bm y})
{\bm N}
\cdot
{\bm F}.
\label{yllls}
\end{align}
where
\begin{align}
{\bm F}&= c_{a_1\cdots a_\ell}({\bm f}_{a_1}\otimes\cdots\otimes{\bm f}_{a_\ell})
\end{align}
is a traceless symmetric $\ell$-tensor,
and
\begin{align}
{\bm N}&=({\bm n}\otimes\cdots\otimes {\bm n})
\end{align}
is an $H$-invariant $\ell$-tensor.
(\ref{yllls}) is a scalar function on ${\bm S}^p$ because
it is invariant under the change of the frame section
$g({\bm y})\rightarrow g'({\bm y})=g({\bm y})h({\bm y})$.

Next, let us consider vector harmonics.
An arbitrary $\RR^{p+1}$-vector function ${\bm S}^p\rightarrow\RR^{p+1}$
that in general has both normal and tangential components to ${\bm S}^p$
can be expanded
by a set of vector functions of the form
\begin{align}
{\bm Y}'^1({\bm y})={\bm f}_b c_{b,a_1a_2\cdots a_\ell}y_{a_1}\cdots y_{a_\ell}.
\label{yprime}
\end{align}
A vector harmonic ${\bm Y}^1:{\bm S}^p\rightarrow T{\bm S}^p$ can be obtained by
projecting away the normal component from (\ref{yprime}).
Its components are given by
\begin{align}
Y_i^1({\bm y})
={\bm\xi}_i^{({\bm y})}\cdot{\bm f}_b c_{b,a_1a_2\cdots a_\ell}y_{a_1}\cdots y_{a_\ell}
=g({\bm y}){\bm N}_i\cdot{\bm F},
\end{align}
where ${\bm F}$ and ${\bm N}_i$ are defined by
\begin{align}
{\bm F}=c_{b,a_1a_2\cdots a_\ell}({\bm f}_b\otimes{\bm f}_{a_1}\otimes\cdots\otimes{\bm f}_{a_\ell}),\quad
{\bm N}_i=({\bm f}_i\otimes{\bm n}\otimes\cdots\otimes{\bm n}).
\end{align}

It is easy to generalize the above construction of the scalar and the vector harmonics
to general spins.
Let $S$ and $R$ be a spin and an angular momentum,
which are representations of $H$ and $G$, respectively.
We suppose these are irreducible.
Let $V_R$ and $\wt V_R$ be the representation space of $R$
and its dual space, respectively.
Although $R$ is an irreducible representation of $G$,
it may be reducible as an $H$ representation.
For the existence of the $S$ harmonics with angular momentum $R$
$S$ must appear in the $H$-irreducible decomposition of $R$.
In other words, there must be an $H$-invariant subspace $V_S\subset V_R$
and its dual space $\wt V_S\subset\wt V_R$ associated with the spin
representation $S$.

Let ${\bm E}_\mu\in V_R$ ($\mu=1,\ldots,\dim R$) be basis vectors of $V_R$,
and $\wt{\bm E}_\mu\in\wt V_R$ be the dual
vectors satisfying $(\wt{\bm E}_\mu,{\bm E}_\nu)=\delta_{\mu\nu}$.
$G$ acts on these vectors as
\begin{align}
g{\bm E}_\mu={\bm E}_\nu\rho_{\nu\mu}^R(g),\quad
g\wt{\bm E}_\mu=\rho_{\mu\nu}^R(g^{-1})\wt{\bm E}_\nu,
\label{basesg}
\end{align}
where $\rho^R_{\mu\nu}(g)=(\wt{\bm E}_\mu,g{\bm E}_\nu)$
is the representation matrix for $R$.
We also introduce basis vectors ${\bm E}_\alpha\in V_S$ and
$\wt{\bm E}_\alpha\in\wt V_S$ that are transformed by $H$ in a similar way to
(\ref{basesg}).

Spin $S$ harmonics with angular momentum $R$
are given by
\begin{align}
Y^{SR}_{\alpha\mu}({\bm y})
=(g({\bm y})\wt{\bm E}_\alpha,{\bm E}_\mu),
\label{genharm}
\end{align}
where $\alpha=1,\ldots,\dim S$ is a spin index,
and $\mu=1,\ldots,\dim R$ labels harmonics belonging to $R$.
The harmonics are transformed
under the local frame rotation
$g'({\bm y})=g({\bm y})h({\bm y})$
by
\begin{align}
Y^{SR}_{\alpha\mu}({\bm y})
&=\rho^S_{\alpha\beta}(h^{-1}({\bm y}))Y^{SR}_{\beta\mu}({\bm y}).
\end{align}
This means $Y_{\alpha\mu}^{SR}$ have spin $S$.
Under an isometry transformation
${\bm y}'=\ol g^{-1}{\bm y}$ by $\ol g\in G$,
$Y^{SR}_{\alpha\mu}$ are transformed as
\begin{align}
Y^{SR}_{\alpha\mu}({\bm y}')
&=\rho^S_{\alpha\beta}(h({\bm y}))Y^{SR}_{\beta\nu}({\bm y})\rho^R_{\nu\mu}(\ol g),
\label{isometryy}
\end{align}
where $h({\bm y})=g^{-1}({\bm y})\ol gg(\ol g^{-1}{\bm y})\in H$
is the local frame rotation compensating the
change of the local frame due to the isometry rotation.
The relation (\ref{isometryy}) shows that the
harmonics have angular momentum $R$.

For a fixed spin $S$ there are infinite number of representations $R$
that contain $S$ in their $H$-irreducible decomposition.
We can show
by using Peter-Weyl theorem
that the collection of
all $Y_{\alpha\mu}^{SR}$ for such representations
form a complete basis of spin $S$ fields.

\subsection{Covariant derivatives}
The harmonics we defined above
are
eigenfunctions of the Laplacian on the sphere.
This is easily shown by expressing
the covariant derivatives
in an algebraic form.

The covariant exterior derivative $D\equiv e^iD_i\equiv d+\wh\omega$ of a
spin-$S$ harmonic $Y_{\alpha\mu}^{SR}=(g({\bm y})\wt{\bm E}_\alpha,{\bm E}_\mu)$ is given by
\begin{align}
DY_{\alpha\mu}^{SR}
&=dY_{\alpha\mu}^{SR}+\rho^S_{\alpha\beta}(\wh\omega)Y_{\beta\mu}^{SR}
\nonumber\\
&
=(dg({\bm y})\wt{\bm E}_\alpha,{\bm E}_\mu)
+\rho^S_{\alpha\beta}(\wh\omega)(g({\bm y})\wt{\bm E}_\beta,{\bm E}_\mu)
\nonumber\\
&
=(g({\bm y})(g^{-1}({\bm y})dg({\bm y})-\wh\omega)\wt{\bm E}_\alpha,{\bm E}_\mu)
\nonumber\\
&
=(g({\bm y})\wh e\wt{\bm E}_\alpha,{\bm E}_\mu).
\end{align}
At the last step we used (\ref{ewformula}).
This is equivalent to
\begin{align}
D_iY_{\alpha\mu}^{SR}
&
=(g({\bm y})\wh T_{i(-1)}\wt{\bm E}_\alpha,{\bm E}_\mu).
\label{dyformula}
\end{align}
Note that
(\ref{dyformula}) is again a harmonic,
and has the general form (\ref{genharm}) of
harmonics with $\wt{\bm E}_\alpha$ replaced by $\wh T_{i(-1)}\wt{\bm E}_\alpha$.
The second derivative of $Y_{\alpha\mu}^{SR}$ is obtained in the same way as
\begin{align}
D_iD_jY_{\alpha\mu}^{SR}
&=(g\wh T_{i(-1)}\wh T_{j(-1)}\wt{\bm E}_\alpha,{\bm E}_\mu).
\label{ddy}
\end{align}
By contracting indices $i$ and $j$, we obtain
\begin{align}
\Delta Y_{\alpha\mu}^{SR}
&=-\lambda Y_{\alpha\mu}^{SR},\quad
\lambda
=-\wh T_{i(-1)}\wh T_{i(-1)}
=C^G_2(R)-C^H_2(S),
\label{eq39}
\end{align}
where $C_2^{SO(n)}(R)$ is the quadratic Casimir of
an $SO(n)$ representation $R$ defined by
\begin{align}
\frac{1}{2}\sum_{a,b=1}^n\rho^R(\wh T_{ab})\rho^R(\wh T_{ab})=-C^{SO(n)}_2(R).
\end{align}
It is also easy to obtain the curvature tensor
$R_{ij}{}^{kl}=\delta_{ik}\delta_{jl}-\delta_{il}\delta_{kj}$
from the anti-symmetric part of (\ref{ddy}).

\section{${\cal N}=1$ superconformal index in 4d}
In this section we demonstrate the calculation of
the superconformal index of vector multiplets
in a 4d ${\cal N}=1$ supersymmetric gauge theory.
The index
is defined in \cite{Romelsberger:2005eg,Kinney:2005ej},
and the relation to short and long multiplets of the superconformal algebra
is investigated.
The superconformal indices for extended supersymmetric theories are
also defined in \cite{Kinney:2005ej}.
In particular, the index for the ${\cal N}=4$ supersymmetric Yang-Mills
theory is calculated, and the agreement in the large $N$ limit
with the corresponding quantity in the gravity dual is confirmed.
The ${\cal N}=1$ superconformal index for a general ${\cal N}=1$
supersymmetric gauge theory is derived in \cite{Romelsberger:2007ec}
using the Lagrangian of theories in ${\bm S}^3\times\RR$ constructed
in \cite{Romelsberger:2005eg}.
The index is used in \cite{Romelsberger:2007ec} to test Seiberg duality
\cite{Seiberg:1994pq}.

There are two kinds of multiplets of 4d ${\cal N}=1$
supersymmetry: vector multiplets and chiral multiplets.
Due to limitations of space
we only consider
vector multiplets.
A vector multiplet consists of a vector field $A_a$,
a gaugino field $\lambda$,
and an auxiliary field $D$.
All component fields belong to the adjoint representation
of the gauge group.

\subsection{${\bm S}^3$ harmonics}
We consider a gauge theory in ${\bm S}^3\times\RR$.
In this section we label coordinates $y^a\in\RR^4$ in a different way from
the previous section.
We use $(y^4,y^1,y^2,y^3)$ instead of $(y^{-1},y^0,y^1,y^2)$.
For ${\bm S}^3\times\RR$ coordinates we use
$x^a$ ($a=1,2,3,4$), $x^i$ ($i=1,2,3$) for ${\bm S}^3$ and
$x^4$ for $\RR$.
The theory has four commuting conserved charges:
the Hamiltonian $H=-\partial/\partial x^4$,
a $U(1)_R$ charge $R$,
and
the angular momenta
$J_1=\frac{1}{i}\wh T_{12}$ and
$J_2=\frac{1}{i}\wh T_{34}$.
The Dirac matrices used in this section are
\begin{align}
\gamma^i=\left(\begin{array}{cc}
0 & \sigma_i \\
\sigma_i & 0
\end{array}\right)\quad
(i=1,2,3),\quad
\gamma^4=\left(\begin{array}{cc}
0 & -i \\
i & 0
\end{array}\right),\quad
\gamma^5=-\gamma^{1234}=\left(\begin{array}{cc}
1 & 0 \\
0 & -1
\end{array}\right).
\label{dirac4d}
\end{align}
The generators $\wh T_{ab}$ ($a,b=1,2,3,4$) of
$G=SO(4)=SU(2)_L\times SU(2)_R$
are related to
$SU(2)_L$ and $SU(2)_R$ generators $\wh T^L_k$ and $\wh T^R_k$ by
\begin{align}
\wh T_{ij}=\epsilon_{ijk}\wh T_k^D=\epsilon_{ijk}(\wh T^L_k+\wh T^R_k),\quad
\wh T_{i4}=\wh T^L_i-\wh T^R_i,
\end{align}
where $\wh T^D_k$ are generators of
the diagonal subgroup $H=SU(2)_D$.

A spin representation $S$ is specified by
$s\in(1/2)\ZZ_{\geq0}$, while
a $G$ representation $R$ is specified by the
left and the right angular momenta $j_L,j_R\in (1/2)\ZZ_{\geq0}$.
The $G$ representation $R=(j_L,j_R)$
must contain the $H$ representation $s$ in its $H$-irreducible
decomposition.
This requires the triangular inequality
\begin{align}
|j_L-j_R|\leq s\leq j_L+j_R.
\end{align}
${\bm S}^3$ harmonics are labeled by six half integers: $Y^{s(j_L,j_R)}_{\alpha(m_L,m_R)}$.
When we do not need to explicitly show $(j_L,j_R)$ and $(m_L,m_R)$,
we often omit them to simplify the expression.
The eigenvalues of the Laplacian are given by
\begin{align}
\Delta Y^s_\alpha=-[2j_L(j_L+1)+2j_R(j_R+1)-s(s+1)]Y^s_\alpha.
\label{lambdas3}
\end{align}
We define
\begin{align}
\rot Y^s_\alpha\equiv -\rho^s_{\alpha\beta}(\wh T_k^D)D_k Y^s_\beta.
\end{align}
We denote the operator by ``$\rot$'' because this becomes the rotation for vector
fields;
$\rot Y^1_i=\epsilon_{ijk}D_j Y^1_k$.
For spinor fields,
this becomes the Dirac operator up to a numerical factor;
$\rot Y^{\frac{1}{2}}_\alpha=-\frac{i}{2}(\sigma_k)_{\alpha\beta}D_k Y^{\frac{1}{2}}_\beta$.
We can calculate eigenvalues of this operator
as
\begin{align}
\rot Y^s_\alpha
&=-\rho^s_{\alpha\beta}(T^D_k)
(g({\bm y})\wh T_{k(-1)}\wt{\bm E}_\beta,{\bm E}_{(m_L,m_R)})
\nonumber\\
&=
(g({\bm y})\wh T_{k(-1)}\wh T^D_k\wt{\bm E}_\alpha,{\bm E}_{(m_L,m_R)}).
\nonumber\\
&=
\sigma Y_\alpha^{\frac{1}{2}},
\label{rotandsigma}
\end{align}
where $\sigma$ is given by
\begin{align}
\sigma=\wh T_{k(-1)}\wh T^D_k
=(\wh T^L_k)^2-(\wh T^R_k)^2
=-j_L(j_L+1)+j_R(j_R+1).
\end{align}

\subsection{Killing spinors}
Supersymmetry transformations in a 4d
conformally flat background
are parameterized by
a left-handed spinor $\epsilon$ and
a right handed spinor $\ol\epsilon$
satisfying
the Killing spinor equations
\begin{align}
D_a\epsilon=\gamma_a\ol\kappa,\quad
D_a\ol\epsilon=\gamma_a\kappa,\quad
(a=1,2,3,4).
\label{killing4d}
\end{align}
To calculate the superconformal index,
we need to determine Killing spinors in ${\bm S}^3\times\RR$.
From (\ref{killing4d})
we can show
$\Delta_{{\bm S}^3} \epsilon=-\frac{3}{4}\epsilon$,
(and the same equation for $\ol\epsilon$).
Comparing this with (\ref{lambdas3}), we find
$\epsilon$ and $\ol\epsilon$ must have
$(j_L,j_R)=(\tfrac{1}{2},0)$ or
$(j_L,j_R)=(0,\tfrac{1}{2})$.
There are four linearly independent Killing spinors for $\epsilon$:
\begin{align}
\epsilon_\alpha({\bm y})\propto
Y_{\alpha(\pm\frac{1}{2},0)}^{\frac{1}{2}(\frac{1}{2},0)},\quad
\epsilon_\alpha({\bm y})\propto
Y_{\alpha(0,\pm\frac{1}{2})}^{\frac{1}{2}(0,\frac{1}{2})}.
\end{align}
We also have four Killing spinors for $\ol\epsilon$.
By using the general formula (\ref{dyformula})
we obtain
\begin{align}
D_i\epsilon
=-\frac{i}{2}\gamma_{\rm iso}\sigma_i\epsilon
=-\frac{1}{2}\gamma_{\rm iso}\gamma_i\gamma_4\epsilon,\quad
D_i\ol\epsilon
=-\frac{i}{2}\gamma_{\rm iso}\sigma_i\ol\epsilon
=\frac{1}{2}\gamma_{\rm iso}\gamma_i\gamma_4\ol\epsilon,
\end{align}
where $\gamma_{\rm iso}$ is the chirality operator for the isometry group $G=SO(4)$,
and $\gamma_{\rm iso}=+1$ ($-1$) for Killing spinors belonging to
$R=(\frac{1}{2},0)$
($R=(0,\frac{1}{2})$).
From these equations, we find
\begin{align}
\ol\kappa=-\frac{1}{2}\gamma_{\rm iso}\gamma_4\epsilon,\quad
\kappa=\frac{1}{2}\gamma_{\rm iso}\gamma_4\ol\epsilon.
\end{align}
The equations in (\ref{killing4d}) with $a=4$ determine the $x^4$ dependence
of the Killing spinors;
\begin{align}
\epsilon\propto e^{-\frac{\gamma_{\rm iso}}{2}x^4},\quad
\ol\epsilon\propto e^{+\frac{\gamma_{\rm iso}}{2}x^4}.
\end{align}

We have obtained eight linearly independent Killing spinors.
($4$ for $\epsilon$ and $4$ for $\ol\epsilon$.)
To perform localization, we choose the following specific ones
\begin{align}
\epsilon_\alpha=e^{-\frac{1}{2}x^4}Y_{\alpha(-\frac{1}{2},0)}^{\frac{1}{2}(\frac{1}{2},0)},\quad
\ol\epsilon_\alpha=e^{\frac{1}{2}x^4}Y_{\alpha(\frac{1}{2},0)}^{\frac{1}{2}(\frac{1}{2},0)},
\end{align}
with quantum numbers
\begin{align}
\epsilon:(H,J_1,J_2,R)=(\tfrac{1}{2},-\tfrac{1}{2},-\tfrac{1}{2},1),\quad
\ol\epsilon:(H,J_1,J_2,R)=(-\tfrac{1}{2},\tfrac{1}{2},\tfrac{1}{2},-1).
\label{eeqn}
\end{align}
We denote the supercharges corresponding to $\epsilon$ and $\ol\epsilon$ by $Q$ and $\ol Q$, respectively.

Among linear combinations of four conserved charges $H$, $J_1$, $J_2$, and $R$,
the following three commute with $Q$ and $\ol Q$.
\begin{align}
\{Q,\ol Q\}=H-J_1-J_2-\frac{3}{2}R,\quad
J_1+\frac{R}{2},\quad
J_2+\frac{R}{2}.
\end{align}
We define the superconformal index by
\begin{align}
I(z_1,z_2)=\tr\left[
(-1)^F
q^{H-J_1-J_2-\frac{3}{2}R}
z_1^{J_1+\frac{R}{2}}
z_2^{J_2+\frac{R}{2}}
\right].
\label{d4indef}
\end{align}

Because the spinors in (\ref{eeqn}) are $SU(2)_R$ singlet,
the cancellation occurs
between modes with the same $SU(2)_R$ quantum numbers.
In the following analysis it is convenient to separate $SU(2)_R$ quantum number from
others.
We define Cartan generators of $SU(2)_L$ and $SU(2)_R$ by
\begin{align}
J_L=\frac{1}{2}(J_1+J_2)=\frac{1}{i}\wh T^L_3,\quad
J_R=\frac{1}{2}(J_1-J_2)=\frac{1}{i}\wh T^R_3,
\end{align}
and rewrite the index
(\ref{d4indef})
by
introducing $z_1=tx$ and $z_2=t/x$ as
\begin{align}
I(z_1,z_2)=\tr\left[
(-1)^F
q^{H-2J_L-\frac{3}{2}R}
t^{2J_L+R}
x^{2J_R}
\right].
\end{align}

\subsection{Mode analysis}
The transformation rules of $4d$ ${\cal N}=1$ vector multiplets are
conformally invariant, and we can use the rules for the flat background
except that the parameters $\epsilon$ and $\ol\epsilon$
are Killing spinors satisfying (\ref{killing4d});
\begin{align}
\delta A_\mu&=i(\epsilon\gamma_\mu\ol\lambda)+i(\ol\epsilon\gamma_\mu\lambda),
\nonumber\\
\delta\lambda&=i\sla F\epsilon+D\epsilon,
\nonumber\\
\delta\ol\lambda&=-i\sla F\ol\epsilon+D\ol\epsilon,
\nonumber\\
\delta D&=(\epsilon\sla D\ol\lambda)+(\ol\epsilon\sla D\lambda),
\end{align}
where $\sla F\equiv (1/2)\gamma^{ab}F_{ab}$ and $\sla D\equiv \gamma^aD_a$.
To perform index calculation, we deform the action by adding $Q$-exact terms.
A standard form of the action used in the literature is $S\sim \delta_Q((\delta_Q\psi)^\dagger\psi)$,
where $\psi$ denotes fermions in the theory.
An advantage of this action is that the bosonic part
$(\delta_Q\psi)^\dagger\delta_Q\psi$ is manifestly positive definite,
and the path integral is well defined automatically.
However, this has also a disadvantage
that it
contains spinor bilinear $\epsilon^\dagger\gamma^i\epsilon$, which
breaks the rotational symmetry $G$.
To avoid this problem, we adopt another action
\begin{align}
{\cal L}
&=\frac{1}{(\ol\epsilon\ol\epsilon')}\delta(\ol\epsilon)\delta(\ol\epsilon')\left(-\frac{1}{4}(\ol\lambda\ol\lambda)\right),
\end{align}
where $\ol\epsilon'$ is
a Killing spinor such that $\ol\epsilon\ol\epsilon'$ is a $G$-invariant scalar.
For example we can use the Killing spinor obtained from $\ol\epsilon$
by replacing 
$Y_{\alpha(\frac{1}{2},0)}^{\frac{1}{2}(\frac{1}{2},0)}$
with $Y_{\alpha(-\frac{1}{2},0)}^{\frac{1}{2}(\frac{1}{2},0)}$ as $\ol\epsilon'$.
The absence of $\ol\epsilon\gamma^i\ol\epsilon'$ in the action is guaranteed by
the algebra $\{\delta(\ol\epsilon),\delta(\ol\epsilon')\}=0$, and
the spinors $\ol\epsilon$ and $\ol\epsilon'$ appear in the action
only through $\ol\epsilon\ol\epsilon'$,
and the Lagrangian is $G$-invariant;
\begin{align}
{\cal L}&=\frac{1}{4}F_{ab}^2-\frac{1}{8}\epsilon^{abcd}F_{ab}F_{cd}
-(\ol\lambda\gamma^a D_a\lambda)-\frac{1}{2}D^2.
\label{eq79}
\end{align}

We fix the gauge symmetry by the gauge fixing function $V_{\rm gf}=D_iA_i$.
The corresponding ghost Lagrangian
is ${\cal L}_{\rm gh}=\ol c'\partial_iD^ic'$, where primes
indicate the ghost and the anti-ghost fields
do not include the constant modes on ${\bm S}^3$.
They are expanded by the scalar harmonics $Y^{0(j,j)}_{(m_L,m_R)}$ ($j\geq\frac{1}{2}$),
and the path integral gives the factor $-4j(j+1)$ for each $j$ and $(m_L,m_R)$.
Note that $V_{\rm gh}$ is
not the full 4d divergence $D_aA_a=D_iA_i+D_4A_4$ but
the 3d divergence in ${\bm S}^3$.
Therefore, the gauge fixing is partial,
and the gauge transformation with parameter
that depends only on $x^4$ is not fixed.
The fixing of this remaining gauge symmetry
has been already discussed in section \ref{intro.sec}, and
gives the Jaccobian factor
in (\ref{qmfix}).

The bosonic part of the Lagrangian including the gauge fixing term is
\begin{align}
{\cal L}
=\frac{1}{2}V_i^2+\frac{1}{2}V_{\rm gf}^2,\quad
V_i=\frac{1}{2}\epsilon_{ijk}F_{jk}-F_{i4},\quad
i,j,k=1,2,3,
\end{align}
where the $D^2$ term is neglected because
the path integral of $D$ gives just a constant.
$(V_i,V_{\rm gf})$ and $A_a$ are related by
\begin{align}
V=D_AA
\end{align}
with a differential operator $D_A$.
The energy eigenmodes are
obtained by solving
$D_AA=0$.
This can be easily solved by harmonic expansion.
We expand $A$ and $V$ as follows.
\begin{align}
&A_i
=a_1Y^{1,(j+1,j)}_{i(m_L,m_R)}
+a_2Y^{1,(j-1,j)}_{i(m_L,m_R)}
+a_3Y^{1,(j,j)}_{i(m_L,m_R)},\quad
A_4=a_4Y^{0,(j,j)}_{(m_L,m_R)},
\nonumber\\
&V_i
=f_1Y^{1,(j+1,j)}_{i(m_L,m_R)}
+f_2Y^{1,(j-1,j)}_{i(m_L,m_R)}
+f_3Y^{1,(j,j)}_{i(m_L,m_R)},\quad
V_{\rm gh}=f_4Y^{0,(j,j)}_{(m_L,m_R)}.
\end{align}

Because $D_A$ is $G$-invariant
modes with different $(j_L,j_R)$ and $(m_L,m_R)$
do not mix.
For $(j_L,j_R)=(j,j)$ ($j\geq\frac{1}{2}$),
there are two components for each $A$ and $V$,
and the coefficients are related by
\begin{align}
\left(\begin{array}{c}
f_3 \\
f_4
\end{array}\right)
&=\left(\begin{array}{cc}
\partial_4 & -1 \\
-4j(j+1) & 0
\end{array}\right)
\left(\begin{array}{c}
a_3 \\
a_4
\end{array}\right)
&
(|m_L|\leq j\geq\frac{1}{2}).
\label{f3f4bya3aa4}
\end{align}
The determinant of the matrix in (\ref{f3f4bya3aa4}) is
\begin{align}
\det=-4j(j+1).
\end{align}
This is canceled by
the ghost factor,
and does not correspond to any physical modes.
When $(j_L,j_R)=(0,0)$, $a_3$ mode does not exist,
and we have only $a_4$ mode.
This corresponds to the Wilson line integrated in (\ref{qmfix}).

The $(j+1,j)$ mode $f_1$ and $(j-1,j)$ mode $f_2$ depend only on
$a_1$ and $a_2$, respectively.
By using formula (\ref{rotandsigma}) we obtain
\begin{align}
f_1&=[\partial_4-2(j+1)]a_1 &(|m_L|\leq j+1),\nonumber\\
f_2&=[\partial_4+2j]a_2 &(|m_L|\leq j-1).
\end{align}
These give the physical modes
in Table \ref{d4amodes.tbl}.
\begin{table}[htb]
\caption{
Bosonic physical modes in ${\bm S}^3\times\RR$ are shown.
We denote $J_L$ eigenvalues by $m$.
$J_R$ eigenvalues always take values between $-j$ and $j$, and we do not
explicitly show them in the table.}\label{d4amodes.tbl}
\centering
\begin{tabular}{ccccc}
\hline
\hline
ID & $H=-\partial_4$ & $D-2J_L-\frac{3}{2}R$ & $2J_L+R$ & range of $m$ \\
\hline
[$A$1] & $-2(j+1)$ & $-2j-2-2m$ &$2m$ & $-j-1\leq m\leq j+1$ \\
{}[$A$2] & $2j$ & $2j-2m$      &$2m$ & $-j+1\leq m\leq j-1$ \\
\hline
\end{tabular}
\end{table}

For the gaugino $\lambda$
we need to solve
the Dirac equation $\sla D\lambda=0$ in ${\bm S}^3\times\RR$.
We expand $\lambda$ and $\sla D\lambda$ by the harmonics
as follows.
\begin{align}
\lambda_\alpha
=cY_{\alpha(m_L,m_R)}^{\frac{1}{2},(j+\frac{1}{2},j)}
+dY_{\alpha(m_L,m_R)}^{\frac{1}{2},(j-\frac{1}{2},j)}.
\quad
\sla D\lambda_\alpha
=sY_{\alpha(m_L,m_R)}^{\frac{1}{2},(j+\frac{1}{2},j)}
+tY_{\alpha(m_L,m_R)}^{\frac{1}{2},(j-\frac{1}{2},j)}
\end{align}
By using (\ref{rotandsigma}) we obtain
\begin{align}
s=-i\left(\partial_4-2j-\frac{3}{2}\right)c,\quad
t=i\left(\partial_4+2j+\frac{1}{2}\right)d.
\end{align}
The corresponding eigenmodes are summarized in Table \ref{lambdamodes}.
\begin{table}[htb]
\caption{
Fermionic physical spectrum in ${\bm S}^3\times\RR$.
We denote $J_L$ eigenvalues by $m'$.
}\label{lambdamodes}
\centering
\begin{tabular}{ccccc}
\hline
\hline
ID & $H=-\partial_4$ & $D-2J_L-\frac{3}{2}R$ & $2J_L+R$ & range of $m'$ \\
\hline
[$\lambda$1] & $-(2j+\frac{3}{2})$ & $-2j-3-2m'$ & $2m'+1$ & $-j-\frac{1}{2}\leq m'\leq j+\frac{1}{2}$ \\
{}[$\lambda$2] & $2j+\frac{1}{2}$ & $2j-1-2m'$ & $2m'+1$ & $-j+\frac{1}{2}\leq m'\leq j-\frac{1}{2}$ \\
\hline
\end{tabular}
\end{table}

Let us compare bosonic modes in Table \ref{d4amodes.tbl} and fermionic modes
in Table \ref{lambdamodes}.
If we replace $m'$ in Table \ref{lambdamodes} by $m-1/2$,
we see that the quantum numbers  $D-2J_L-\frac{3}{2}R$ and $2J_L+R$
in the two tables completely match.
However, the ranges of $m$ are different.
$m=m'+1/2$ runs over the range $-j\leq m\leq j+1$ for [$\lambda$1] and
$-j+1\leq m\leq j$ for [$\lambda$2].
This means 
[$A$1] mode with $m=-j-1$ and
[$\lambda$2] mode with $m=m'+1/2=j$
do not have partners and make contributions to the index.
The [$A$1] modes with $m=-j-1$
contribute to the index by
\begin{align}
I_{\rm sp}(z_1,z_2,e^{ia_I})
=\sum_\alpha\sum_{j\in\ZZ_{\geq0}} e^{-i\alpha(a)}t^{2j+2}\chi_j(x^2)
=\frac{t^2}{(1-tx)(1-t/x)}\sum_\alpha e^{-i\alpha(a)},
\label{s3acont}
\end{align}
where $\chi_j(x^2)$ is the $SU(2)$ character $\chi_j(x^2)=x^{2j}+\cdots+x^{-2j}$.
The [$\lambda$2] modes with $m'=j-1/2$
contribute to the index by
\begin{align}
I^{({\rm tot})}_{\rm sp}(z_1,z_2,e^{ia_I})=-\sum_\alpha\sum_{j\in\frac{1}{2}\ZZ_{\geq1}}e^{i\alpha(a)}t^{2j}\chi_j(x^2)
=\left(1-\frac{1}{(1-tx)(1-t/x)}\right)\sum_\alpha e^{i\alpha(a)}.
\label{s3lcont}
\end{align}
Combining 
the bosonic contribution (\ref{s3acont}),
the fermionic contribution (\ref{s3lcont}),
and the ghost contribution (\ref{ispghost}),
we obtain the total single-particle index
\begin{align}
I^{({\rm tot})}_{\rm sp}(z_1,z_2,e^{ia_I})
&=
\sum_{\alpha}e^{i\alpha(a)}
\left(\delta_{\alpha,0}-\frac{1-z_1z_2}{(1-z_1)(1-z_2)}\right).
\end{align}

Refer to \cite{Romelsberger:2007ec} 
for the result for chiral multiplets.

Before ending this section, we comment on a close relation between 
the 4d superconformal index and the ${\bm S}^3$ partition function
of 3d supersymmetric field theories.
The 3d partition function
of ${\cal N}=2$ supersymmetric field theories
on round ${\bm S}^3$
is first calculated in \cite{Kapustin:2009kz}
for canonical fields.
It is generalized to non-canonical fields in \cite{Jafferis:2010un,Hama:2010av},
and squashed ${\bm S}^3$ in \cite{Hama:2011ea}.
(See also \cite{Imamura:2011wg}.)
Once we obtain the formula for the 4d superconformal index, it is easy to obtain
a general formula for the ${\bm S}^3$ partition function by taking
a small radius limit $\beta\rightarrow 0$
\cite{Dolan:2011rp,Gadde:2011ia,Imamura:2011uw}.

\section{${\cal N}=2$ superconformal index in 3d}
In this section we consider a 3d ${\cal N}=2$ supersymmetric theory.
3d superconformal index is defined in
\cite{Bhattacharya:2008zy}, and the relation to multiplets
of the superconformal algebra is investigated.
An important application of the index is a test of
AdS$_4$/CFT$_3$ correspondence.
For example, the index of the ABJM model \cite{Aharony:2008ug}
is calculated in \cite{Bhattacharya:2008bja}
for the perturbative sector and the agreement with the
corresponding quantity on the gravity side
is confirmed.
The check including monopole contribution is done in
\cite{Kim:2009wb}, in which
formula of the ${\cal N}=2$ superconformal index for canonical fields (without anomalous dimensions) is derived.
The index for chiral multiplets with anomalous dimensions is derived in \cite{Imamura:2011su}.

Just like the 4d ${\cal N}=1$ case, we have two kinds of multiplets:
vector multiplets and chiral multiplets.
Again we only consider a vector multiplet,
which consists of
a vector field $A_a$, a scalar field $\sigma$,
a gaugino field $\lambda$, and an auxiliary field $D$.

We use different labeling of the coordinates $y^a$ from section \ref{harm.sec}.
Instead of $(y^{-1},y^0,y^1)$, we use $(y^3,y^1,y^2)$.
For ${\bm S}^2\times\RR$
we use coordinates $x^a$ ($a=1,2,3$),
$x^i$ ($i=1,2$) for ${\bm S}^2$ and $x^3$ for $\RR$.
We use Pauli matrices as the Dirac matrices;
$\gamma_a=\sigma_a$.

We consider an ${\cal N}=2$ supersymmetric theory in ${\bm S}^2\times\RR$.
There are three conserved charges:
the Hamiltonian $H=-\partial/\partial x^3$,
the angular momentum $J_3=-i\wh T_{12}$,
and a $U(1)_R$ charge $R$.

An important difference of ${\bm S}^2$ from ${\bm S}^3$
discussed in the previous section is
the existence of topologically
non-trivial gauge field configurations,
monopoles.
Harmonics describing charged particles
coupling to such monopole backgrounds
are called monopole harmonics.

\subsection{Monopole harmonics}

Let us consider harmonics with angular momentum $j\in\frac{1}{2}\ZZ_{\geq0}$.
The representation space $V_j$ is spanned by
\begin{align}
{\bm E}_s,\quad
s=-j,-j+1,\ldots,j-1,j,
\end{align}
satisfying
\begin{align}
\wh T_{12}{\bm E}_s=is{\bm E}_s.
\end{align}
The dual vectors are defined by
$\wt{\bm E}_s
={\bm E}_{-s}
=({\bm E}_s)^*$.
All representations of $H$ are singlets.
Spin $s$ harmonics are given by
\begin{align}
Y_{sm}^j
=(g({\bm y})\wt{\bm E}_s,{\bm E}_m)
=\rho^j_{s,m}(g({\bm y})),
\end{align}
where $\rho^j$ is the $SU(2)$ representation matrix for angular momentum $j$.

Let us consider a field with
$U(1)$ electric charge $q\in\ZZ$ in the monopole background
with magnetic charge $m\in\ZZ$.
We can take the gauge
so that the gauge potential is given by
$A=(m/2)\omega_{12}$,
and then the covariant derivative for a spin $s$ field becomes
\begin{align}
D=d+is\omega_{12}-iqA
=d+i\left(s-\frac{mq}{2}\right)\omega_{12}.
\end{align}
Therefore, we can treat a charged spin $s$ particle in the magnetic flux
as a particle with shifted spin
\begin{align}
s_{\rm eff}=s-\frac{qm}{2}.
\label{seff}
\end{align}
Thus we do not have to introduce anything new
for dealing with monopole backgrounds.
In the case of non-Abelian gauge theory,
$m=\sum_Im_It_I$ becomes a Cartan element of
the gauge algebra, and
the factor $qm$ in (\ref{seff}) must be replaced
by $\alpha(m)$ for the components of an adjoint field
specified by the weight $\alpha$.

It is convenient to define $v_\pm$ for a general vector ${\bm v}$
by $v_\pm\equiv v_1\pm iv_2$.
For example, we define
\begin{align}
D_\pm\equiv D_1\pm iD_2,\quad
\wh T_{3\pm}\equiv \wh T_{31}\pm i\wh T_{32}.
\end{align}
The basis ${\bm E}_s$ with $s=\pm1$ are given by
${\bm E}_{\pm1}={\bm f}_1\pm i{\bm f}_2$.
We should note that
spin $\pm1$ components of a vector ${\bm v}$
are
$\wt E_{\pm1}\cdot{\bm v}=v_\mp$.
The representation matrices
$(T_{3\pm})_{ss'}\equiv \rho^j_{ss'}(\wh T_{3\pm})$ have non-vanishing components
$(\wh T_{3\pm})_{s\pm1,s}$,
and satisfy
\begin{align}
(T_{3\pm})_{s,s\mp1}(T_{3\mp})_{s\mp1,s}&=-[j(j+1)-s(s\mp1)]=-(j\pm s)(j\mp s+1).
\end{align}
From the general formula (\ref{dyformula}) and (\ref{eq39}) we obtain
\begin{align}
D_\pm Y_{s,m}^j
=(T_{3\pm})_{s,s\mp1}Y_{s\mp1,m}^j,\quad
\Delta Y_{s,m}^j&=-[j(j+1)-s^2]Y_{s,m}^j.
\end{align}

\subsection{Killing spinors}
The parameters of ${\cal N}=2$ supersymmetry are two-component
spinors $\epsilon$ and $\ol\epsilon$.
In a Euclidean space they are treated as independent spinors.
They must satisfy the Killing spinor equations
$D_\mu\epsilon=\gamma_\mu\ol\kappa$ and
$D_\mu\ol\epsilon=\gamma_\mu\kappa$.
We can easily show that each component of $\epsilon$ and $\ol\epsilon$
must be a spinor harmonic with $j=\frac{1}{2}$.
A general form of such spinors are
\begin{align}
\epsilon
=\left(\begin{array}{c}
Y_{\frac{1}{2},m}^{\frac{1}{2}}c_m(x^3) \\
Y_{-\frac{1}{2},m}^{\frac{1}{2}}c'_m(x^3)
\end{array}\right),\quad
\ol\epsilon
=\left(\begin{array}{c}
Y_{\frac{1}{2},m}^{\frac{1}{2}}\ol c_m(x^3) \\
Y_{-\frac{1}{2},m}^{\frac{1}{2}}\ol c'_m(x^3)
\end{array}\right),
\label{s2rkilling}
\end{align}
where $m$ are summed over $\pm\frac{1}{2}$.
We can determine the $x^3$ dependence of the coefficients
in the same way as the 4d case,
and obtain eight linearly independent Killing spinors.
We use the following specific ones for localization.
\begin{align}
\epsilon
=e^{-\frac{1}{2}x^3}
\left(\begin{array}{c}
Y_{\frac{1}{2},-\frac{1}{2}}^{\frac{1}{2}} \\
Y_{-\frac{1}{2},-\frac{1}{2}}^{\frac{1}{2}}
\end{array}\right),\quad
\ol\epsilon
=e^{+\frac{1}{2}x^3}
\left(\begin{array}{c}
Y_{\frac{1}{2},\frac{1}{2}}^{\frac{1}{2}} \\
-Y_{-\frac{1}{2},\frac{1}{2}}^{\frac{1}{2}}
\end{array}\right).
\end{align}
$\epsilon$ and $\ol\epsilon$ carry $(H,J_3,R)=\pm(\frac{1}{2},-\frac{1}{2},1)$.
We denote the corresponding supercharges by $Q$ and $\ol Q$.

Among linear combinations of the three conserved charges $H$, $J_3$, and $R$,
the following two commute with $Q$ and $\ol Q$.
\begin{align}
\{Q,\ol Q\}=H-J_3-R,\quad
2J_3+R.
\end{align}
We define the superconformal index by
\begin{align}
I(x,m_I)=\tr\left[(-1)^F
q^{H-J_3-R}
x^{2J_3+R}
\right],
\end{align}
where $m_I$ are the background monopole charges.

\subsection{Mode analysis}
The supersymmetry transformation rules for
3d ${\cal N}=2$ vector multiplets
in a conformally flat background are
\begin{align}
\delta A_a&=i(\epsilon\gamma_a\ol\lambda)-i(\ol\epsilon\gamma_a\lambda),
\nonumber\\
\delta\sigma&=(\epsilon\ol\lambda)+(\ol\epsilon\lambda),
\nonumber\\
\delta\lambda&=i\sla F\epsilon-(\sla D\sigma)\epsilon-2\sigma\ol\kappa+D\epsilon,
\nonumber\\
\delta\ol\lambda&=-i\sla F\ol\epsilon-(\sla D\sigma)\ol\epsilon-2\sigma\kappa+D\ol\epsilon,
\nonumber\\
\delta D&=-(\epsilon\sla D\ol\lambda)-(\ol\epsilon\sla D\lambda)
+(\ol\kappa\ol\lambda)+(\kappa\lambda)
+(\epsilon[\sigma,\ol\lambda])
-(\ol\epsilon[\sigma,\lambda]).
\end{align}

For localization we use
the $\ol Q$-exact Lagrangian
\begin{align}
{\cal L}
&=\frac{1}{(\ol\epsilon\ol\epsilon')}
\delta(\ol\epsilon)\delta(\ol\epsilon')
\left(\frac{1}{4}\ol\lambda\ol\lambda\right)
\nonumber\\
&=\frac{1}{4}F_{ab}F^{ab}
-\frac{1}{2}\epsilon^{abc}F_{ab}D_c\sigma
+\frac{1}{2}D_a\sigma D^a\sigma
+\frac{1}{2}\sigma^2
-\sigma F_{12}
+\sigma D^3\sigma
\nonumber\\
&-(\ol\lambda\gamma^a D_a\lambda)
-\ol\lambda[\sigma,\lambda]
-\frac{1}{2}(\ol\lambda\gamma^3\lambda)
-\frac{1}{2}D^2,
\end{align}
where $\ol\epsilon'$ is the Killing spinor obtained from $\ol\epsilon$ by
replacing
$Y_{s,\frac{1}{2}}^{\frac{1}{2}}$
by $Y_{s,-\frac{1}{2}}^{\frac{1}{2}}$.

The Lagrangian for $A$ and $\sigma$ is rewritten in the
manifestly positive definite form ${\cal L}=(1/2)V_aV_a$
with
\begin{align}
V_a=D_a\sigma+\delta_{a3}\sigma-\frac{1}{2}\epsilon_{abc}F_{bc},
\end{align}
and saddle points are given by $V_a=0$.
This condition is solved,
up to gauge transformation,
by the monopole backgrounds
\begin{align}
A=A^{(0)}\equiv \frac{m}{2}\omega_{12},\quad
\sigma=\sigma^{(0)}\equiv \frac{m}{2},\quad
m=\sum_Im_It_I,
\label{bkgs2}
\end{align}
where $m\in\ZZ$ is the monopole charge.
These backgrounds induce the shift of the effective spin
\begin{align}
s_{\rm eff}=s+s_0,\quad
s_0=-\frac{\alpha(m)}{2}=-\alpha(\sigma_0),
\end{align}
for a field associated with a weight $\alpha$.
We consider fluctuations around the saddle points.
\begin{align}
A_a=A_a^{(0)}+\delta A_a,\quad
\sigma=\sigma^{(0)}+\delta\sigma.
\end{align}
We fix the gauge symmetry by
the gauge fixing function
$V_{\rm gf}
=D_i^{(0)}\delta A_i
\equiv\partial_i\delta A_i-i[A_i^{(0)},\delta A_i]$.
The quadratic part of the Lagrangian including the gauge fixing term is
\begin{align}
{\cal L}_{\rm bos}=\frac{1}{2}V_aV_a+\frac{1}{2}V_{\rm gf}V_{\rm gf}.
\end{align}

We expand the bosonic fields by the
monopole harmonics as follows.
\begin{align}
&
\delta A_+=Y_{s_0-1,m}^j a_+,\quad
\delta A_-=Y_{s_0+1,m}^j a_-,\quad
\delta A_3=Y_{s_0,m}^j a_3,\quad
\delta \sigma=Y_{s_0,m}^j a_4,\nonumber\\
&
V_+=Y_{s_0-1,m}^j a_+,\quad
V_-=Y_{s_0+1,m}^j a_-,\quad
V_3=Y_{s_0,m}^j a_3,\quad
V_{\rm gh}=Y_{s_0,m}^j v_4.
\end{align}
Unlike the 4d case,
harmonics in the four components all have the same
$G$ quantum numbers $j$ and $m$,
and can mix among them.
When $j\geq|s_0|+1$,
all four components exist,
while if $j$ is equal to or smaller than $|s_0|$
some of them are absent.
When all components exist the coefficients are
related by
\begin{align}
\left(\begin{array}{c}
v_+ \\ v_- \\ v_3 \\ v_4
\end{array}\right)
=
\left(
\begin{array}{cccc}
-is_0-i\partial_3 & 0 & 
+i(T_{3+})_{s,s-1} &
(T_{3+})_{s,s-1} \\
0 & -is_0+i\partial_3 &
-i(T_{3-})_{s,s+1} & (T_{3-})_{s,s+1} \\
\frac{i}{2}(T_{3-})_{s-1,s} &
-\frac{i}{2}(T_{3+})_{s+1,s} &
-is_0 & \partial_3+1 \\
\frac{1}{2}(T_{3-})_{s-1,s} &
\frac{1}{2}(T_{3+})_{s+1,s} &
\partial_3 & 0
\end{array}\right)
\left(\begin{array}{c}
a_+ \\ a_- \\ a_3 \\ a_4
\end{array}\right),
\end{align}
and the determinant of the $4\times4$ matrix is
\begin{align}
\det=[j(j+1)-s_0^2](\partial_3-j)(\partial_3+j+1),\quad j\geq|s_0|+1.
\end{align}
For smaller $j$ some of rows and columns are absent, and
the determinant is given by
\begin{align}
\det&=\pm i(D_3+j+1),\quad(j=|s_0|-1),\nonumber\\
\det&=\pm i(j(j+1)-s_0^2)(D_3+j+1),\quad(j=|s_0|).
\end{align}

The factor $[j(j+1)-s_0^2]$ is canceled by the factor arising
from
the ghost term
${\cal L}_{\rm gh}=\ol c'D_i^{(0)}D_ic'$
and does not correspond to any physical modes.
Other two factors, $\partial_3-j$ and $\partial_3+j+1$ correspond to physical modes
in Table \ref{amodes3d}.
\begin{table}[htb]
\caption{Bosonic physical spectrum in ${\bm S}^2\times\RR$.
We denote $J_3$ eigenvalues by $m$.}\label{amodes3d}
\centering
\begin{tabular}{cccccc}
\hline
\hline
ID & $H=-\partial_3$ &range of $j$ & $D-J_3-R$ & $2J_3+R$ & range of $m$ \\
\hline
{}[$A$1] & $-j$   & $|s_0|+1\leq j$ & $-j-m$ & $2m$ & $-j\leq m\leq j$ \\
{}[$A$2] & $j+1$ & $|s_0|-1\leq j$ & $j+1-m$ & $2m$ & $-j\leq m\leq j$ \\
\hline
\end{tabular}
\end{table}

The fermionic part of the $\ol Q$-exact Lagrangian is
${\cal L}=\ol\lambda D_\lambda\lambda$ with
\begin{align}
D_\lambda=-\gamma^a D_a-\sigma-\frac{1}{2}\gamma^3
=\left(\begin{array}{cc}
-\partial_3-\sigma-\frac{1}{2} & -D_- \\
-D_+ & \partial_3-\sigma+\frac{1}{2} \\
\end{array}\right)
\end{align}
We expand the components of $\lambda$
and $D_\lambda\lambda$ by
\begin{align}
\lambda=\left(\begin{array}{c}
c Y_{s_0+\frac{1}{2},m}^j \\
d Y_{s_0-\frac{1}{2},m}^j
\end{array}\right),\quad
D_\lambda\lambda=\left(\begin{array}{c}
s Y_{s_0+\frac{1}{2},m}^j \\
t Y_{s_0-\frac{1}{2},m}^j
\end{array}\right).
\end{align}
When $j\geq|s_0|+\frac{1}{2}$,
the two components
are non vanishing, and
the relation among the coefficients is
\begin{align}
\left(\begin{array}{c}
s \\ t
\end{array}\right)
=
\left(\begin{array}{cc}
-\partial_3+s-\frac{1}{2} & -(T_{3-})_{s-\frac{1}{2},s+\frac{1}{2}} \\
-(T_{3+})_{s+\frac{1}{2},s-\frac{1}{2}} & \partial_3+s+\frac{1}{2} \\
\end{array}\right)
\left(\begin{array}{c}
c \\ d
\end{array}\right).
\end{align}
The determinant of the matrix is
\begin{align}
\det=-(\partial_3+j+1)(\partial_3-j),\quad |s_0|+\frac{1}{2}\leq j.
\end{align}
When $j=|s_0|-\frac{1}{2}$
only one of the two components of $\lambda$ and $D_\lambda\lambda$
exists, and the matrix element becomes $\pm(\partial_3+j+1)$.
The physical modes are summarized in Table \ref{d3lmodes}.
\begin{table}[htb]
\caption{Fermionic physical spectrum in ${\bm S}^2\times\RR$.
For convenience we use $j'$ and $j''$ instead of $j$ and $m'$ instead of $m$.
}\label{d3lmodes}
\centering
\begin{tabular}{cccccc}
\hline
\hline
ID & $H=-\partial_3$ &range of $j$ & $D-J_3-R$ & $2J_3+R$ & range of $m'$ \\
\hline
{}[$\lambda$1] & $-j'$   & $|s_0|+\frac{1}{2}\leq j'$ & $-j'-1-m'$ & $2m'+1$ & $-j'\leq m'\leq j'$ \\
{}[$\lambda$2] & $j''+1$ & $|s_0|-\frac{1}{2}\leq j''$ & $j''-m'$ & $2m'+1$ & $-j''\leq m'\leq j''$ \\
\hline
\end{tabular}
\end{table}

Let us compare the bosonic and fermionic modes.
By the replacement
\begin{align}
j'=j-\frac{1}{2},\quad
j''=j+\frac{1}{2},\quad
m'=m-\frac{1}{2},
\end{align}
the ranges of $j$, $D-J_3-R$, and $2J_3+R$ are completely
match between the two tables.
However, the ranges of $m$ are different.
The range is $-j+1\leq m\leq j$ for [$\lambda$1],
and
$-j\leq m\leq j+1$ for [$\lambda$2].
Therefore, [$A$1] with $m=-j$
and [$\lambda$2] with $m=j+1$, ($m'=j''$) contribute
to the index.
The former
contribute
\begin{align}
I_{\rm sp}
=\sum_\alpha\sum_{j=|s_0|+1}^\infty e^{-i\alpha(a)}x^{2j}
=\sum_\alpha e^{-i\alpha(a)}\frac{x^{|\alpha(m)|+2}}{1-x^2},
\label{ispa3d}
\end{align}
and the latter
contribute
\begin{align}
I_{\rm sp}
=\sum_\alpha\sum_{j=|s_0|-1}^\infty e^{i\alpha(a)}x^{2j+2}
=\sum_\alpha e^{i\alpha(a)}\left(-\frac{x^{|\alpha(m)|}}{1-x^2}+\delta_{\alpha(m),0}
\right),
\label{ispl3d}
\end{align}
to the index.

For the ghost index (\ref{ispghost}),
we must take account of the gauge symmetry breaking
due to the monopole background.
The ghost zero-modes are present only for unbroken symmetry,
and the ghost index becomes
\begin{align}
I_{\rm sp}^{({\rm gh})}
=\sum_{\alpha\neq0,\alpha(m)=0}e^{-i\alpha(a)}
=\sum_{\alpha}e^{i\alpha(a)}(\delta_{\alpha,0}-\delta_{\alpha(m),0}).
\label{monopilgh}
\end{align}

By combining
(\ref{ispa3d}),
(\ref{ispl3d}),
and (\ref{monopilgh}),
we obtain
\begin{align}
I^{({\rm tot})}_{\rm sp}(x,e^{ia_I},m_I)
=\sum_\alpha e^{i\alpha(a)}
\left(
\delta_{\alpha,0}
-x^{|\alpha(m)|}
\right).
\end{align}

Refer to \cite{Imamura:2011su} for the chiral multiplet
contribution.

\section{${\cal N}=(1,0)$ superconformal index in 6d}\label{d6sc}
The superconformal indices in 6d theories
are defined in \cite{Bhattacharya:2008zy}.
The derivation by harmonic expansion in this section is based on \cite{Imamura:2012bm}.
The method used in \cite{Imamura:2012bm} is essentially the same as
that in \cite{Kim:2012av},
in which
the ${\bm S}^5$ partition function is calculated
by using
$\CP^2$ harmonics in \cite{Pope:1980ub}.
See also \cite{Kallen:2012cs,Kallen:2012va,Lockhart:2012vp}
for the ${\bm S}^5$ partition function.

Although it is known that instantons make non-perturbative
contribution to the index,
it seems in practice impossible to calculate it by harmonic expansion,
and
we are going to calculate only perturbative contribution to the index.

\subsection{Hopf fibration}\label{sec51}
In this section we consider a 6d ${\cal N}=(1,0)$ supersymmetric theory
in ${\bm S}^5\times\RR$.
As in the 3d and 4d cases,
we need to choose a particular supercharge
for localization.
The choice breaks the
rotational symmetry $G=SO(6)$.
Unlike the 3d and 4d cases
it seems impossible to construct $Q$-exact Lagrangian
that respect the full $SO(6)$ symmetry.
The Lagrangian we use respects
only the subgroup $G'=SU(3)\times U(1)$.
Fortunately,
this is transitive,
and harmonic expansion is still efficient.
Because of this symmetry breaking,
it is natural to regard ${\bm S}^5$ as
a Hopf fibration over $\CP^2$.
In this subsection we discuss
the Hopf fibration of general odd-dimensional spheres,
and define a coordinate system convenient
for the following analysis.

Let us consider ${\bm S}^{2r+1}$ with unit radius defined
as the subset of $\RR^{2r+2}$ by $y_ay_a=1$.
In this subsection
we use indices $a,b,\ldots=-1,0,1,\ldots,2r$,
$i,j,\ldots=0,1,\ldots,2r$,
and $m,n,\ldots=1,\ldots,2r$.
The first step to define the Hopf fibration is to specify a
complex structure in $\RR^{2r+2}$.
Let $\II_{ab}$ be the complex structure
in $\RR^{2r+2}$ with non-vanishing components
\begin{align}
\II_{(-1)0}=\II_{12}=\cdots=\II_{(2r-1)(2r)}=1,
\end{align}
and
let $U(1)_\II$ be the subgroup of $G\equiv SO(2r+2)$
generated by $\wh\II=(1/2)\II_{ab}\wh T_{ab}$.
We define Hopf fibration with $U(1)_\II$ orbits.
${\bm y}\in{\bm S}^{2r+1}$ can be written as
\begin{align}
{\bm y}(\theta,\psi)=e^{-\psi\wh\II}{\bm y}_0(\theta),
\label{hopfy}
\end{align}
where $0\leq\psi<2\pi$ is a coordinate along fibers, and
${\bm y}_0(\theta)$ is a representative in the
fiber that is specified by $\CP^2$ coordinates
$\theta=(\theta^1,\theta^2,\theta^3,\theta^4)$.
The ${\bm S}^{2r+1}$ metric is
\begin{align}
|d{\bm y}|^2
&
=(d\psi+V)^2
+ds_{\CP^r}^2,
\label{s5andcp2metric}
\end{align}
where $V$ and $ds_{\CP^r}^2$ are defined by
\begin{align}
V
=-d{\bm y}_0\cdot \wh\II{\bm y}_0
=\II_{ab}y_0^ady_0^b,\quad
ds_{\CP^2}^2=|d{\bm y}_0|^2
-(d{\bm y}_0\cdot \wh\II{\bm y}_0)^2.
\end{align}
$ds_{\CP^r}^2$ is the Fubini-Study metric of $\CP^r$.

A convenient choice of the local frame is
given by a frame section
of the form
\begin{align}
g(\theta,\psi)=e^{-\psi\wh\II}g_0(\theta),\quad
g_0(\theta)\in SU(r+1),
\label{unitaryframe}
\end{align}
where $SU(r+1)$ is the special unitary group that
rotates holomorphic vectors ${\bm f}_{2k-1}+i{\bm f}_{2k}$
($k=0,\ldots,r$).
We call this a unitary frame.
Because
$\partial_\psi{\bm y}(\theta,\psi)={\bm\xi}_0^{({\bm y})}$
the $0$ direction points the fiber direction.
The vielbein
$e^0$ is given by
\begin{align}
e^0=d\psi+V,
\label{ezero}
\end{align}
and the others $e^m$ ($m=1,2,\ldots,2r$) are the pull-back of the
vielbein of
the base $\CP^r$.

The exterior derivative of $V$ is
\begin{align}
dV
=\II_{ab}dy_0^a\wedge dy_0^b
=\II_{ab}dy^a\wedge dy^b.
\label{dv1}
\end{align}
At the second equality we use $|{\bm y}|=1$.
Clearly $dV$ satisfies
$(\partial_\psi,dV)=0$, and
can be regarded as the pull-back of a two-form in the base $\CP^r$.
The unitary transformation $g(\theta,\psi)$ keeps
the complex structure intact,
and
the components of $dV$ in (\ref{dv1}) in the unitary frame
is essentially the same as $\II_{ab}$
except that $(-1)$ direction is absent in ${\bm S}^{2r+1}$.
Namely, $dV$ is given by
\begin{align}
dV=I_{mn}e^m\wedge e^n,
\end{align}
where $I_{mn}$ is an anti-symmetric $2r\times2r$ matrix with non-vanishing components
\begin{align}
I_{12}=I_{34}=\cdots=I_{(2r-1)(2r)}=1.
\end{align}
This is nothing but the complex structure in the
base $\CP^r$.

From the torsionless condition, we obtain
the spin connection
\begin{align}
\wh\omega\equiv \frac{1}{2}\omega_{ij}\wh T_{ij}=-e^0\wh I+e^mI_{mn}\wh T_{n0}+\wh\omega^{\CP^r},
\label{omegas5cpe}
\end{align}
where $\wh\omega^{\CP^r}\equiv(1/2)\omega_{mn}^{\CP^r}\wh T_{mn}$ is the spin connection
of $\CP^r$.
Let us read off the vielbein $e^i$ and the spin connection $\omega_{ij}$
according to (\ref{ewformula})
from the Maurer-Cartan form
\begin{align}
g^{-1}dg=g_0^{-1}dg_0-d\psi\wh\II.
\end{align}
Because $g\in U(r+1)$, $SO(2r+2)$ generators $\wh T_{i(-1)}$, whose coefficients
are identified with the vielbein $e^i$, always appear in the
Maurer-Cartan form through the combinations
\begin{align}
\wh K_0=\wh T_{0(-1)}+\frac{1}{r}\wh I,\quad
\wh K_m=\wh T_{m(-1)}+I_{mn}\wh T_{n0},\quad(m=1,2,\ldots,2r).
\end{align}
Therefore, the components of $\omega_{ij}$ corresponding to $\wh I$ and $\wh T_{n0}$
are written in terms of $e^i$.
We obtain
\begin{align}
\wh\omega=-e^0\wh I
+e^mI_{mn}\wh T_{n0}
+\wh\omega^{SU(r)}
+\frac{r+1}{r}V\wh I,
\label{uiing}
\end{align}
where $\wh\omega^{SU(r)}$ is the $SU(r)$ part of the spin connection
cting on the holomorphic vectors ${\bm f}_{2k-1}+i{\bm f}_{2k}$
($k=1,\ldots,r$).
From (\ref{omegas5cpe}) and (\ref{uiing}),
we obtain
\begin{align}
\wh\omega^{\CP^r}=\wh\omega^{SU(r)}+\frac{r+1}{r}V\wh I.
\end{align}

Under the subgroup $G'$
a $G$ representation $R$ is decomposed into
$\oplus_i(R_i',q_i)$
where $R_i'$ and $q_i$ are $SU(r+1)$ representations and
$U(1)_\II$ charges.
Correspondingly,
${\bm S}^{2r+1}$ harmonics are
expanded into
Kaluza-Klein modes by
\begin{align}
Y^R(\theta,\psi)=\sum_ie^{iq_i\psi}Y^{(R_i,q_i)}(\theta).
\end{align}
By applying the covariant derivative to this expansion,
we obtain
\begin{align}
DY^R(\theta,\psi)=\sum_i
e^{iq_i\psi}(\nabla+e^mI_{mn}\wh T_{n0}+e^0(iq_i-\wh I))Y^{(R_i,q_i)}(\theta).
\end{align}
$\nabla$ is the covariant derivative on $\CP^r$ defined by
\begin{align}
\nabla
=d-iQ_VV+\wh\omega^{\CP^r}
=d-iQ_VV+\frac{r+1}{r}V\wh I+\wh\omega^{SU(r)},
\label{nabladef}
\end{align}
where $Q_V\equiv \frac{1}{i}\wh\II$.

\subsection{Killing spinors}
Let us construct Killing spinors in ${\bm S}^5\times\RR$.
We use $6,5,1,2,3,4$ instead of $-1,0,1,2,3,4$ used in \ref{sec51}.
Namely, $1234$, $5$, and $6$ label $\CP^2$, Hopf fibers, and Euclidean time,
respectively.
We use the 6d Dirac matrices
\begin{align}
\Gamma_7=\left(\begin{array}{cc}
1 \\
& -1
\end{array}\right),\quad
\Gamma_6=\left(\begin{array}{cc}
& -i \\
i
\end{array}\right),\quad
\Gamma_i
=\left(\begin{array}{cc}
& \gamma_i \\
\gamma_i
\end{array}\right),
\end{align}
where $\gamma^i$ are the five dimensional Dirac matrices
defined in (\ref{dirac4d}).
The charge conjugation matrix in 6d and 5d are
\begin{align}
C^{(6)}
=\left(\begin{array}{cc}
& C^{(5)} \\
-C^{(5)}
\end{array}\right),\quad
C^{(5)}
=\left(\begin{array}{cc}
\epsilon \\
& -\epsilon
\end{array}\right).
\end{align}

The supersymmetry transformation parameter $\epsilon^I$ of ${\cal N}=(1,0)$
supersymmetry in an arbitrary conformally flat background
is a left-handed symplectic Majorana-Weyl spinor satisfying
the Killing spinor equation
\begin{equation}
D_a\epsilon^I=\Gamma_a\kappa^I,\quad
a=1,\ldots,6.
\label{6dkilling}
\end{equation}
$\kappa^I$ is a right-handed symplectic Majorana-Weyl spinor.
The R-symmetry is $SU(2)_R$, and
both $\epsilon^I$ and $\kappa^I$ have $SU(2)_R$ doublet index $I=1,2$.

To write down the Killing spinors in ${\bm S}^5\times\RR$,
we first define Killing spinors in ${\bm S}^5$.
There are two spinor representations of $G=SO(6)$.
Let $V^{(L)}$ and $V^{(R)}$ be representation spaces for positive and negative chirality, respectively.
We introduce basis vectors ${\bm E}^{(L)}_\mu$ ($\mu=1,2,3,4$) for $V^{(L)}$ and
${\bm E}^{(R)}_{\dot\mu}$ ($\dot\mu=\dot1,\dot2,\dot3,\dot4$) for $V^{(R)}$.
Both these spinor representations are irreducible also as $H$-representations,
and we can identify $V_S$ with $V^{(L)}$ and $V^{(R)}$.
According to general prescription,
we can define
$8$ linearly independent Killing spinors on ${\bm S}^5$:
\begin{align}
\varepsilon^{(L)}_{\alpha\mu}=(g({\bm y})\wt{\bm E}_\alpha,{\bm E}_\mu^{(L)}),\quad
\varepsilon^{(R)}_{\alpha\dot\mu}=(g({\bm y})\wt{\bm E}_\alpha,{\bm E}_{\dot\mu}^{(R)}),
\end{align}
where $\wt{\bm E}_\alpha$ are dual basis vectors for $V_S$.
By using the general formula (\ref{dyformula}) and the
explicit representation of the Dirac matrices,
we obtain
\begin{align}
D_i\varepsilon^{(L/R)}=-\frac{i}{2}\Gamma_{\rm iso}\gamma_i\varepsilon^{(L/R)},
\quad (i=1,\ldots,5),
\label{deps}
\end{align}
where $\Gamma_{\rm iso}$
is the chirality operator of the isometry group $G=SO(6)$.
Under the subgroup
$G'=SU(3)\times U(1)_\II$,
each of
$\varepsilon^{(L)}_{\alpha\mu}$ and
$\varepsilon^{(R)}_{\alpha\dot\mu}$
splits into a singlet and a triplet.
This can be seen from the explicit form of
$\wh\II$ for the spinor representations:
\begin{align}
\rho^{(L)}(\wh\II)=\frac{1}{2}\diag(i,-3i,i,i),\quad
\rho^{(R)}(\wh\II)=\frac{1}{2}\diag(3i,-i,-i,-i).
\end{align}
We can see that
$\varepsilon^{(L)}_{\alpha\mu}$ and
$\varepsilon^{(R)}_{\alpha\dot\mu}$
belong respectively to
the following $G'$ representations.
\begin{align}
\varepsilon^{(L)}_{\alpha\mu}:
(\ol{\bm 3},+\frac{1}{2})
\oplus({\bm 1},-\frac{3}{2}),\quad
\varepsilon^{(R)}_{\alpha\dot\mu}:
({\bm 3},-\frac{1}{2})
\oplus({\bm 1},+\frac{3}{2}).
\end{align}
For the index calculation we use the $SU(3)$ singlet Killing spinors
\begin{align}
\varepsilon^1\equiv
\varepsilon_{\alpha\dot 1}^{(R)}
=e^{+\frac{3i}{2}\psi}\delta_{\alpha1},\quad
\varepsilon^2\equiv
\varepsilon_{\alpha2}^{(L)}
=e^{-\frac{3i}{2}\psi}\delta_{\alpha2}.
\end{align}
We can show
that these $\varepsilon^I$ are $\nabla$-constant.
\begin{align}
\nabla\varepsilon^I
=0.
\end{align}
This property drastically simplify the following calculation.

We define the 6d Killing spinors $\epsilon^I$ in ${\bm S}^5\times\RR$
by using ${\bm S}^5$ Killing spinors $\varepsilon^I$ by
\begin{align}
\epsilon^I=
e^{-\frac{1}{2}\Gamma_{\rm iso}x^6}
\left(
\begin{array}{c}
\varepsilon^I \\
0
\end{array}\right),\quad
\kappa^I=-\frac{1}{2}\Gamma_{\rm iso}\Gamma_6\varepsilon^I.
\end{align}

There are five conserved charges:
the Hamiltonian $H=-\partial_6$,
the angular momenta
$J_1=-i\wh T_{12}$,
$J_2=-i\wh T_{34}$,
$J_3=-i\wh T_{65}$,
and the $SU(2)_R$ Cartan generator $R_3$.
The supercharges $Q_I$ corresponding to $\epsilon^I$
carry charges
$(H,J_1,J_2,J_3,R_3)=\pm(\frac{1}{2},-\frac{1}{2},-\frac{1}{2},-\frac{1}{2},1)$.
The following four commute with $Q\equiv Q_1+Q_2$.
\begin{align}
Q^2
=H-J_1-J_2-J_3-2R_3,\quad
J_1+\frac{R_3}{2},\quad
J_2+\frac{R_3}{2},\quad
J_3+\frac{R_3}{2}.
\end{align}
We define the superconformal index
\begin{align}
I(z_1,z_2,z_3)=\tr\left[(-1)^F
q^{H-J_1-J_2-J_3-2R_3}
z_1^{J_1+\frac{R_3}{2}}
z_2^{J_2+\frac{R_3}{2}}
z_3^{J_3+\frac{R_3}{2}}
\right].
\end{align}

Because the supercharges $Q_I$ are $SU(3)$ singlets,
we can treat the $SU(3)$ as a ``flavor'' symmetry, and
it is convenient to separate $SU(3)$ fugacities from $z_i$.
We use variables
$x$ and $y_i$ defined by
$z_i=xy_i$ ($y_1y_2y_3=1$) instead of $z_i$.
The index is rewritten in terms of these variables by
\begin{align}
I(z_1,z_2,z_3)=\tr\left[(-1)^F
q^{H-Q_V-2R_3}
x^{Q_V+\frac{3}{2}R_3}
y^{SU(3)}
\right],
\end{align}
where $Q_V=J_1+J_2+J_3=\frac{1}{i}\wh\II$ and
$y^{SU(3)}\equiv y_1^{J_1} y_2^{J_2} y_3^{J_3}$.

\subsection{$\CP^2$ harmonics}
According to the general construction in section \ref{harm.sec},
the scalar spherical harmonics in ${\bm S}^5$
are characterized by rank $k$ symmetric traceless tensors in $\RR^6$.
Let $R_k$ denote the tensor representation.
Under $G'=SU(3)\times U(1)$
$R_k$ is decomposed into $k+1$ representations $R_{k,q}$ ($q=k,k-2,\ldots,-k$)
where $R_{k,q}$ is the symmetric traceless part of the product
$({\bm3},+1)^{\frac{k+q}{2}}\otimes(\ol{\bm3},-1)^{\frac{k-q}{2}}$.
Let ${\bm E}^{(k,q)}_\mu$ ($\mu=1,\ldots,\dim R_{k,q}$) be basis vectors of $V_{R_{k,q}}$.
By definition ${\bm E}^{(k,q)}_\mu$ satisfy
$Q_V{\bm E}^{(k,q)}_\mu=q{\bm E}^{(k,q)}_\mu$.
The corresponding scalar harmonics are given by
\begin{align}
Y^{0(k,q)}_\mu({\bm y})=(g({\bm y}){\bm N},{\bm E}^{(k,q)}_\mu).
\end{align}
The following relations hold.
\begin{align}
\partial_\psi Y^{0(k,q)}_\mu&=iqY^{0(k,q)}_\mu,\nonumber\\
\nabla_m\nabla_m Y^{0(k,q)}_\mu&=-\lambda_{k,q}Y^{0(k,q)}_\mu,\quad
\lambda_{k,q}=k(k+4)-q^2,\nonumber\\
[\nabla_m,\nabla_n]Y^{0(k,q)}_\mu&=-2iqI_{mn}Y^{0(k,q)}_\mu.
\label{nanay}
\end{align}
We normalize $Y^{0(k,q)}_\mu$ by
\begin{align}
\int_{{\bm S}^5} |Y^{0(k,q)}_\mu|^2=1.
\end{align}
For simplicity of expressions
we omit the $G'$ indices $\mu$ and $(k,q)$
in the following.


By using the scalar harmonics and the Killing spinors $\varepsilon^I$
we can define
a complete set of
spinor harmonics
\begin{align}
Y^{\frac{1}{2}(1)}=\varepsilon^{(R)}Y^0,\quad
Y^{\frac{1}{2}(2)}=\varepsilon^{(L)}Y^0,\quad
Y^{\frac{1}{2}(3)}=\gamma^m\varepsilon^{(R)}\nabla_m Y^0,\quad
Y^{\frac{1}{2}(4)}=\gamma^m\varepsilon^{(L)}\nabla_m Y^0.
\end{align}
Spinor harmonics with different $G'$ quantum numbers are orthogonal.
The inner products of two harmonics with the same $G'$ quantum numbers
are
\begin{equation}
G^{\frac{1}{2}}_{AB}=\int_{{\bm S}^5} (Y^{\frac{1}{2}(A)})^\dagger Y^{\frac{1}{2}(B)}=\left(\begin{array}{cccc}
1 & 0 & 0 & 0 \\
0 & 1 & 0 & 0 \\
0 & 0 & \lambda_{k,q}-4q & 0 \\
0 & 0 & 0 & \lambda_{k,q}+4q
\end{array}\right).
\end{equation}
Among four harmonics $Y^{\frac{1}{2}(1)},\ldots,Y^{\frac{1}{2}(4)}$,
some may not exist when $k$ and $q$ take boundary values.
For example, the norm of $Y^{\frac{1}{2}(3)}$,
$|Y^{\frac{1}{2}(3)}|^2=k(k+4)-q(q+4)$
vanishes when $q=k$, and then $Y^{\frac{1}{2}(3)}$ does not exists.
Similarly, $Y^{\frac{1}{2}(4)}$ does not exist when $q=-k$,
and
only $Y^{\frac{1}{2}(1)}$ and $Y^{\frac{1}{2}(2)}$ exist
when $k=q=0$.

A complete set of vector harmonics can be defined by
\begin{equation}
Y_m^{1(A)}=X_{mn}^{(A)}\nabla_n Y^0\quad
(A=1,2,3,4),
\end{equation}
where $X_{mn}^{(A)}$ are the tensors with the components
\begin{align}
X^{(a)}_{mn}=\epsilon_{amn}+(\delta_{am}\delta_{4n}-\delta_{an}\delta_{4m}),\quad
X^{(4)}_{mn}=\delta_{mn}.
\end{align}
$X^{(a)}_{mn}$ ($a=1,2,3$) can be defined also as the following bi-linears
of the Killing spinors.
\begin{align}
&X_{mn}^{(3)}
=I_{mn}
=-i(\varepsilon^{1\dagger}\gamma_{mn}\varepsilon^1)
=i(\varepsilon^{2\dagger}\gamma_{mn}\varepsilon^2),\nonumber\\
&X^{(1)}_{mn}+iX^{(2)}_{mn}
=-i(\varepsilon^{2\dagger}\gamma_{mn}\varepsilon^1),\nonumber\\
&X^{(1)}_{mn}-iX^{(2)}_{mn}
=-i(\varepsilon^{1\dagger}\gamma_{mn}\varepsilon^2).
\end{align}
These are $\nabla$-constant.
The inner products of two vector harmonics with the same $G'$
quantum numbers are
\begin{equation}
G^1_{AB}
=\int (Y_m^{1(A)})^* Y_m^{1(B)}
=\left(\begin{array}{cccc}
\lambda_{k,q} & 4iq \\
-4iq & \lambda_{k,q} \\
& & \lambda_{k,q} & -4iq \\
& & 4iq & \lambda_{k,q}
\end{array}\right).
\end{equation}

\subsection{Mode analysis}
A vector multiplet consists of
a vector field $A_a$, a gaugino $\lambda_I$ and a $SU(3)$ triplet auxiliary field $D_{IJ}$.
The transformation laws are
\begin{align}
\delta A_a&=-(\epsilon^I\Gamma_a\lambda_I),\nonumber\\
\delta \lambda_I&=-\sla F\epsilon_I+iD_I{}^J\epsilon_J,\nonumber\\
\delta D_{IJ}&=2i(\epsilon_{\{I}\sla D\lambda_{J\}})-4i(\kappa_{\{I}\lambda_{J\}}).
\end{align}
For index calculation
we use the $Q$-exact action
\begin{align}
{\cal L}
=&Q[(Q\lambda_I)^\dagger\lambda_I]
\nonumber\\
=&\frac{1}{2}F_{mn}^2+\frac{1}{4}\epsilon^{mnpq}F_{mn}F_{pq}
+F_{56}^2+F_{5m}^2+F_{6m}^2-\frac{1}{2}D_I{}^JD_J{}^I
\nonumber\\&
+\lambda^I\left(
-\gamma^i D_i
-iD_6
-\frac{1}{4}I_{mn}\gamma^{mn}-\frac{i}{2}\tau_3\gamma^5+\frac{i}{2}\tau_3\right)\lambda_I,
\label{defaction6}
\end{align}
where $\tau_3=\diag(1,-1)$ is the matrix acting on $SU(2)_R$ doublets.

The auxiliary fields $D_{IJ}$ does not give any physical modes.

The gauge field terms in (\ref{defaction6})
with the gauge fixing term $(D_i A^i)^2$ added
can be rewritten as
\begin{align}
{\cal L}_V
&=\frac{1}{2}\sum_{a=1}^3(X_{mn}^{(a)} F_{mn})^2
+F_{56}^2+F_{5m}^2+F_{6m}^2+(D_i A^i)^2.
\label{vac}
\end{align}
For the saddle points all terms in 
(\ref{vac}) must vanish.
For the first term to vanish
the gauge field in $\CP^2$ must be anti-self-dual.
This allows instanton configurations.
However,
it is difficult to calculate the
instanton contribution to the index by means of harmonic expansion.
Here we focus only on the perturbative sector
with zero instanton number.
Then the saddle point is given by the trivial gauge configuration
$A_a=0$ up to gauge transformations.
Let us represent the fluctuation of the gauge potential as a column vector ${\cal A}$
and expand it by six basis vectors ${\cal Y}_{1,\ldots,6}$.
Their explicit forms are
\begin{align}
{\cal A}=\left(\begin{array}{c}
A_m \\
A_5 \\
A_6
\end{array}\right),\quad
{\cal Y}_{1,2,3,4}=
\left(\begin{array}{c}
Y_m^{1(1,2,3,4)} \\
0 \\
0
\end{array}\right),\quad
{\cal Y}_5=
\left(\begin{array}{c}
0 \\
Y^0 \\
0
\end{array}\right),\quad
{\cal Y}_6=
\left(\begin{array}{c}
0 \\
0 \\
Y^0
\end{array}\right).
\end{align}
We can write the Lagrangian in the form ${\cal L}={\cal A}^\dagger{\cal D}{\cal A}$
with a certain differential operator ${\cal D}$.
It is straightforward to calculate the determinant
of the differential operator ${\cal D}$ in each subspace
with specific $G'$ quantum numbers $(k,q)$ and $\mu$.
For $|q|\leq k-2$, all six basis vectors are linearly independent,
and the determinant is
\begin{align}
\det{\cal D}
&=\frac{\det\int{\cal Y}_A^\dagger{\cal D}{\cal Y}_B}{\det\int{\cal Y}_A^\dagger{\cal Y}_B}
\nonumber\\
&=k^2(k+4)^2
(k^2-\partial_6^2)
((k+4)^2-\partial_6^2)
\nonumber\\&\quad
\times
(k^2+4k+9-2q-\partial_6^2)
(k^2+4k+9+2q-\partial_6^2).
\label{vdetg}
\end{align}
The factor $k^2(k+4)^2$ is canceled by the ghost factor,
and the other factors correspond to physical modes.
For $|q|=k$, the six vectors ${\cal Y}_{1,\ldots,6}$ are not linearly independent,
and some factors in 
(\ref{vdetg}) are absent.
Careful analysis gives the spectrum in Table \ref{table:vectmodes}.
\begin{table}[htb]
\caption{Bosonic physical modes in the representation $R_{k,q}$.
$\sqrt{\pm}=\sqrt{k^2+4k+9\pm 2q}$.}
\label{table:vectmodes}
\begin{center}
\begin{tabular}{ccllc}
\hline
\hline
ID & $H=-\partial_6$    & $H-Q_V-2R$ & $Q_V+\frac{3}{2}R_3$ & range of $q$ \\
\hline
{}[$A1$] & $k+4$    & $k+4-q$         & $q$   & $-k\leq q\leq k$ \\
{}[$A2$] & $k$        & $k-q$           & $q$   & $-k+2\leq q\leq k-2$ \\
{}[$A3$] & $\sqrt{+}$ & $\sqrt{+}-q-3$  & $q+3$ & $-k\leq q\leq k-2$ \\
{}[$A4$] & $\sqrt{-}$ & $\sqrt{-}-q+3$  & $q-3$ & $-k+2\leq q\leq k$ \\
\hline
{}[$\ol{A1}$] & $-(k+4)$    & $-k-4-q$        & $q$   & $-k\leq q\leq k$ \\
{}[$\ol{A2}$] & $-k$        & $-k-q$          & $q$   & $-k+2\leq q\leq k-2$ \\
{}[$\ol{A3}$] & $-\sqrt{-}$ & $-\sqrt{-}-q+3$ & $q-3$ & $-k+2\leq q\leq k$ \\
{}[$\ol{A4}$] & $-\sqrt{+}$ & $-\sqrt{+}-q-3$ & $q+3$ & $-k\leq q\leq k-2$ \\
\hline
\end{tabular}
\end{center}
\end{table}


The gaugino terms in (\ref{defaction6}) take the form
$\lambda^I D_\lambda\lambda_I$ with the differential operator
\begin{align}
D_\lambda=
-\gamma^i D_i
-iD_6
-\frac{1}{4}I_{mn}\gamma^{mn}-\frac{i}{2}\tau_3\gamma^5+\frac{i}{2}\tau_3
\end{align}
We define matrix representation $M_{AB}$ of this operator by
${\cal D}_\lambda Y^{\frac{1}{2}(A)}=Y^{\frac{1}{2}(B)} M_{BA}$.
For $|q|\leq k-2$
we obtain
\begin{equation}
M_{AB}=\left(\begin{array}{cccc}
-i(q+\frac{7}{2}+\partial_6) & 0 & (\lambda_{k,q}-4q) & 0 \\
0 & -i(q-\frac{7}{2}+\partial_6) & 0 & (\lambda_{k,q}+4q) \\
-1 & 0 & i(q+\frac{3}{2}-\partial_6+\tau_3) & 0 \\
0 & -1 & 0 & i(q-\frac{3}{2}-\partial_6+\tau_3)
\end{array}\right).
\end{equation}
and the determinant is
\begin{align}
\det M_{ij}&=
\left[\partial_6-\tau_3\left(k+\frac{7}{2}\right)\right]
\left[\partial_6+\tau_3\left(k+\frac{1}{2}\right)\right]
\left[\left(\partial_6+\frac{1}{2}\tau_3\right)^2-(k^2+4k+9+2\tau_3q)\right].
\end{align}
Again we have to analyze $|q|=k$ case separately.
The physical spectrum is shown in Table \ref{table:s5rgauginomodes}.
\begin{table}[htb]
\caption{Fermionic physical modes in the representation $R_{k,q}$.
$\sqrt{\pm}=\sqrt{k^2+4k+9\pm 2q}$.
}
\label{table:s5rgauginomodes}
\begin{center}
\begin{tabular}{ccllcc}
\hline
\hline
ID & $R_3$ & $H=-\partial_6$ & $H-Q_V-2R_3$ & $Q_V+\frac{3}{2}R_3$ & range of $q$ \\
\hline
{}[$\lambda1$] & $-1$ & $-\frac{1}{2}+(k+4)$  & $(k+4)-q$  & $q$      & $-k\leq q\leq k$ \\
{}[$\lambda2$] & $+1$ & $\frac{1}{2}+k$      & $k-q$      & $q$      & $-k+2\leq q\leq k$ \\
{}[$\lambda3$] & $+1$ & $\frac{1}{2}+\sqrt+$ & $\sqrt+-q-3$ & $q+3$  & $-k\leq q\leq k$ \\
{}[$\lambda4$] & $-1$ & $-\frac{1}{2}+\sqrt-$ & $\sqrt--q+3$ & $q-3$ & $-k+2\leq q\leq k$ \\
\hline
{}[$\ol{\lambda1}$] & $+1$ & $\frac{1}{2}-(k+4)$  & $-(k+4)-q$    & $q$   & $-k\leq q\leq k$ \\
{}[$\ol{\lambda2}$] & $-1$ & $-\frac{1}{2}-k$      & $-k-q$        & $q$   & $-k\leq q\leq k-2$ \\
{}[$\ol{\lambda3}$] & $-1$ & $-\frac{1}{2}-\sqrt-$ & $-\sqrt--q+3$ & $q-3$ & $-k\leq q\leq k$ \\
{}[$\ol{\lambda4}$] & $+1$ & $\frac{1}{2}-\sqrt+$ & $-\sqrt+-q-3$ & $q+3$ & $-k\leq q\leq k-2$ \\
\hline
\end{tabular}
\end{center}
\end{table}

Comparing
Table \ref{table:vectmodes} and
Table \ref{table:s5rgauginomodes},
all quantum numbers match except the ranges of $q$.
Let us focus on the positive frequency modes
because $A_a$ are real and $\lambda_I$ are symplectic Majorana.
Only difference of the bosonic and the fermionic spectra is that
there are $q=k$ modes in [$\lambda2$] and [$\lambda3$]
but not in [$A2$] and [$A3$].
The modes [$\lambda2$] and [$\lambda3$] with $q=k$
contribute to the index by
\begin{align}
I_{\rm sp}(z_i,e^{ia_I})
=&\tr\left[(-1)^Fq^{H-Q_V-2R_3}x^{Q_V+\frac{3}{2}R_3}y^{SU(3)}e^{ia_It_I}\right]
\nonumber\\
=&
-\sum_\alpha\sum_{k=1}^\infty e^{i\alpha}x^k\chi_{(k,k)}(y)
-\sum_\alpha\sum_{k=0}^\infty e^{i\alpha}x^{k+3}\chi_{(k,k)}(y),
\end{align}
where $\chi_{(k,q)}(y)$ is the $SU(3)$ character of the representation $R_{k.q}$.
By combining this with (\ref{ispghost}) we obtain
\begin{align}
I^{({\rm tot})}_{\rm sp}(z_i,e^{ia_I})
=\sum_\alpha e^{i\alpha(a)}\left(\delta_{\alpha,0}-\frac{1+z_1z_2z_3}{(1-z_1)(1-z_2)(1-z_3)}\right),
\end{align}
where we used the formula
\begin{align}
\sum_{k=0}^\infty x^k\chi_{(k,k)}(y)
=\frac{1}{(1-z_1)(1-z_2)(1-z_3)},
\end{align}
for the $SU(3)$ characters.

Refer to \cite{Imamura:2012bm}
for index calculation for hypermultiplets
by means of harmonic expansion.

\section{Concluding Remarks}
We reviewed how we calculate
3d, 4d, and 6d
superconformal indices
by using supersymmetric localization and
harmonic expansion.
We deformed the theories
by introducing $Q$-exact terms,
and derived the indices by means of
mode analysis in ${\bm S}^p\times\RR$
backgrounds.

The method of harmonic expansion works effectively
when the deformed theory has $G=SO(p+1)$ rotational symmetry.
This is the case in 3d and 4d.
Although it is not the case in 6d and the deformed Lagrangian
respect only the subgroup $G'=U(3)\in SO(6)$,
$G'$ is still transitive and
harmonic expansion is usufull.

In the case of 5d, however, this method does not work.
Let $Q$ be a supercharge that we use for localization.
In general $Q^2$ is a linear combination of a $G$ generator $T$
and generators of internal symmetries.
Unlike the ${\bm S}^5$ case we discussed in section \ref{d6sc},
in which $T$ generates shifts along Hopf fibers,
${\bm S}^4$ rotations generated by $T$ always have (at least) two fixed points.
These fixed points are often called North and South poles,
and the existence of such special points
clearly shows that the symmetry of the deformed theory
is not transitive on ${\bm S}^4$.
For this reason it is not practical to use
harmonic expansion
for the index calculation.
This is also the case for the ${\bm S}^4$ partition function of
4d theories.

The 5d superconformal index and the ${\bm S}^4$ partition function
have been calculated by using more sophisticated technique,
called equivariant localization.
The ${\bm S}^4$ partition function is calculated in the seminal work by Pestun \cite{Pestun:2007rz},
and the result is extended to squashed ${\bm S}^4$ in \cite{Hama:2012bg}.
The 5d superconformal index is derived in \cite{Kim:2012gu,Terashima:2012ra}.
All these works employ the method of equivariant localization.

In the method using equivariant localization, the existence of fixed points
is not a trouble but what is required.
The partition function and the index are given as the product of
the contribution of each fixed point.
This also make it possible to includes
the instanton contribution as the contribution from fixed points,
each of which is given by the Nekrasov partition function
\cite{Nekrasov:2002qd,Nekrasov:2003rj}.

Unfortunately, we have no space to discuss these issues, and
the interested reader is reffered to the original works
cited above.

\section*{Acknowledgment}
This work was partially supported by Grand-in-Aid for Scientific Research (C) (No.15K05044),
Ministry of Education, Science and Culture, Japan.


\begin{thebibliography}{99}

\bibitem{Witten:1982df} 
  E.~Witten,
  ``Constraints on Supersymmetry Breaking,''
  Nucl.\ Phys.\ B {\bf 202}, 253 (1982).
\bibitem{Romelsberger:2005eg} 
  C.~Romelsberger,
  ``Counting chiral primaries in N = 1, d=4 superconformal field theories,''
  Nucl.\ Phys.\ B {\bf 747}, 329 (2006)
  [hep-th/0510060].
\bibitem{Kinney:2005ej}
  J.~Kinney, J.~M.~Maldacena, S.~Minwalla and S.~Raju,
  ``An Index for 4 dimensional super conformal theories,''
  Commun.\ Math.\ Phys.\  {\bf 275}, 209 (2007)
  [hep-th/0510251].
\bibitem{Bhattacharya:2008zy} 
  J.~Bhattacharya, S.~Bhattacharyya, S.~Minwalla and S.~Raju,
  ``Indices for Superconformal Field Theories in 3,5 and 6 Dimensions,''
  JHEP {\bf 0802}, 064 (2008)
  [arXiv:0801.1435 [hep-th]].
\bibitem{Romelsberger:2007ec} 
  C.~Romelsberger,
  ``Calculating the Superconformal Index and Seiberg Duality,''
  arXiv:0707.3702 [hep-th].
\bibitem{Seiberg:1994pq} 
  N.~Seiberg,
  ``Electric - magnetic duality in supersymmetric nonAbelian gauge theories,''
  Nucl.\ Phys.\ B {\bf 435}, 129 (1995)
  [hep-th/9411149].



\bibitem{Kapustin:2009kz} 
  A.~Kapustin, B.~Willett and I.~Yaakov,
  ``Exact Results for Wilson Loops in Superconformal Chern-Simons Theories with Matter,''
  JHEP {\bf 1003}, 089 (2010)
  [arXiv:0909.4559 [hep-th]].

\bibitem{Jafferis:2010un} 
  D.~L.~Jafferis,
  ``The Exact Superconformal R-Symmetry Extremizes Z,''
  JHEP {\bf 1205}, 159 (2012)
  [arXiv:1012.3210 [hep-th]].

\bibitem{Hama:2010av} 
  N.~Hama, K.~Hosomichi and S.~Lee,
  ``Notes on SUSY Gauge Theories on Three-Sphere,''
  JHEP {\bf 1103}, 127 (2011)
  [arXiv:1012.3512 [hep-th]].


\bibitem{Hama:2011ea} 
  N.~Hama, K.~Hosomichi and S.~Lee,
  ``SUSY Gauge Theories on Squashed Three-Spheres,''
  JHEP {\bf 1105}, 014 (2011)  [arXiv:1102.4716 [hep-th]].

\bibitem{Imamura:2011wg} 
  Y.~Imamura and D.~Yokoyama,
  ``N=2 supersymmetric theories on squashed three-sphere,''
  Phys.\ Rev.\ D {\bf 85}, 025015 (2012)
  [arXiv:1109.4734 [hep-th]].

\bibitem{Dolan:2011rp} 
  F.~A.~H.~Dolan, V.~P.~Spiridonov and G.~S.~Vartanov,
  ``From 4d superconformal indices to 3d partition functions,''
  Phys.\ Lett.\ B {\bf 704}, 234 (2011)
  [arXiv:1104.1787 [hep-th]].
\bibitem{Gadde:2011ia}
  A.~Gadde and W.~Yan,
  ``Reducing the 4d Index to the $S^3$ Partition Function,''
  arXiv:1104.2592 [hep-th].
\bibitem{Imamura:2011uw}
  Y.~Imamura,
  ``Relation between the 4d superconformal index and the S$^3$ partition function,''
  [arXiv:1104.4482 [hep-th]].

\bibitem{Aharony:2008ug} 
  O.~Aharony, O.~Bergman, D.~L.~Jafferis and J.~Maldacena,
  ``N=6 superconformal Chern-Simons-matter theories, M2-branes and their gravity duals,''
  JHEP {\bf 0810}, 091 (2008)
  [arXiv:0806.1218 [hep-th]].


\bibitem{Bhattacharya:2008bja} 
  J.~Bhattacharya and S.~Minwalla,
  ``Superconformal Indices for N = 6 Chern Simons Theories,''
  JHEP {\bf 0901}, 014 (2009)
  [arXiv:0806.3251 [hep-th]].


\bibitem{Kim:2009wb} 
  S.~Kim,
  ``The Complete superconformal index for N=6 Chern-Simons theory,''
  Nucl.\ Phys.\ B {\bf 821}, 241 (2009)
  [Nucl.\ Phys.\ B {\bf 864}, 884 (2012)]
  [arXiv:0903.4172 [hep-th]].

\bibitem{Imamura:2011su} 
  Y.~Imamura and S.~Yokoyama,
  ``Index for three dimensional superconformal field theories with general R-charge assignments,''
  JHEP {\bf 1104}, 007 (2011)
  [arXiv:1101.0557 [hep-th]].








\bibitem{Imamura:2012bm} 
  Y.~Imamura,
  ``Perturbative partition function for squashed S$^5$,''
  arXiv:1210.6308 [hep-th].

\bibitem{Kim:2012av} 
  H.~-C.~Kim and S.~Kim,
  ``M5-branes from gauge theories on the 5-sphere,''
  arXiv:1206.6339 [hep-th].

\bibitem{Pope:1980ub} 
  C.~N.~Pope,
  ``EIGENFUNCTIONS AND SPIN (c) STRUCTURES IN CP**2,''
  Phys.\ Lett.\ B {\bf 97}, 417 (1980).

\bibitem{Kallen:2012cs} 
  J.~Kallen and M.~Zabzine,
  ``Twisted supersymmetric 5D Yang-Mills theory and contact geometry,''
  JHEP {\bf 1205}, 125 (2012)
  [arXiv:1202.1956 [hep-th]].
\bibitem{Kallen:2012va} 
  J.~Kallen, J.~Qiu and M.~Zabzine,
  ``The perturbative partition function of supersymmetric 5D Yang-Mills theory with matter on the five-sphere,''
  arXiv:1206.6008 [hep-th].
\bibitem{Lockhart:2012vp} 
  G.~Lockhart and C.~Vafa,
  ``Superconformal Partition Functions and Non-perturbative Topological Strings,''
  arXiv:1210.5909 [hep-th].

\bibitem{Kim:2012qf} 
  H.~C.~Kim, J.~Kim and S.~Kim,
  ``Instantons on the 5-sphere and M5-branes,''
  arXiv:1211.0144 [hep-th].


\bibitem{Pestun:2007rz} 
  V.~Pestun,
  ``Localization of gauge theory on a four-sphere and supersymmetric Wilson loops,''
  Commun.\ Math.\ Phys.\  {\bf 313}, 71 (2012)  [arXiv:0712.2824 [hep-th]].

\bibitem{Hama:2012bg} 
  N.~Hama and K.~Hosomichi,
  ``Seiberg-Witten Theories on Ellipsoids,''
  JHEP {\bf 1209}, 033 (2012)
  [JHEP {\bf 1210}, 051 (2012)]
  [arXiv:1206.6359 [hep-th]].

\bibitem{Kim:2012gu} 
  H.~C.~Kim, S.~S.~Kim and K.~Lee,
  ``5-dim Superconformal Index with Enhanced En Global Symmetry,''
  JHEP {\bf 1210}, 142 (2012)
  [arXiv:1206.6781 [hep-th]].
\bibitem{Terashima:2012ra} 
  S.~Terashima,
  ``Supersymmetric gauge theories on $S^4$ x $S^1$,''
  Phys.\ Rev.\ D {\bf 89}, no. 12, 125001 (2014)
  [arXiv:1207.2163 [hep-th]].


\bibitem{Nekrasov:2002qd} 
  N.~A.~Nekrasov,
  ``Seiberg-Witten prepotential from instanton counting,''
  Adv.\ Theor.\ Math.\ Phys.\  {\bf 7}, 831 (2004)
  [hep-th/0206161].

\bibitem{Nekrasov:2003rj} 
  N.~Nekrasov and A.~Okounkov,
  ``Seiberg-Witten theory and random partitions,''
  hep-th/0306238.
\end{thebibliography}
\end{document}